\documentclass[suppldata]{interact}
\usepackage{fix-cm}
\usepackage{epstopdf}
\usepackage{graphicx}
\usepackage{float}
\usepackage{footnote}
\usepackage{xfrac}
\usepackage{here}
\usepackage{url}
\usepackage{subcaption} 
\usepackage[margin=10pt,font={sf,small},labelfont=sf]{caption}
\usepackage{color}
\usepackage{natbib}
\usepackage{algorithm}
\usepackage{algpseudocode}
\usepackage{etoolbox}

\theoremstyle{plain}

\theoremstyle{definition}

\theoremstyle{remark}

\newcommand{\be}{\begin{equation}}
\newcommand{\ee}{\end{equation}}

\newcommand{\mc}[1]{\mathcal{#1}}

\makeatletter

\newcommand*{\algrule}[1][\algorithmicindent]{
  \makebox[#1][l]{
    \hspace*{.2em}
    \vrule height .75\baselineskip depth .25\baselineskip
  }
}

\newcount\ALG@printindent@tempcnta
\def\ALG@printindent{
    \ifnum \theALG@nested>0
    \ifx\ALG@text\ALG@x@notext
    \else
    \unskip
    \ALG@printindent@tempcnta=1
    \loop
    \algrule[\csname ALG@ind@\the\ALG@printindent@tempcnta\endcsname]
    \advance \ALG@printindent@tempcnta 1
    \ifnum \ALG@printindent@tempcnta<\numexpr\theALG@nested+1\relax
    \repeat
    \fi
    \fi
}
\patchcmd{\ALG@doentity}{\noindent\hskip\ALG@tlm}{\ALG@printindent}{}{\errmessage{failed to patch}}
\patchcmd{\ALG@doentity}{\item[]\nointerlineskip}{}{}{}
\makeatother

\begin{document}

\title{\begin{center} 
Practical Implementation of an End-to-End Methodology for SPC of 3-D Part Geometry: A Case Study \\
 \end{center}}

\author{
\begin{center}
\name{Yulin An\textsuperscript{$\ast$}\thanks{Y.A. e-mail: yba5115@psu.edu}, Xueqi Zhao\textsuperscript{$\dagger$}\thanks{X.Z. e-mail xuz206.psu@gmail.com} and Enrique del Castillo\textsuperscript{$\ast,\ddagger$}\thanks{E.D.C., corresponding author, e-mail: exd13@psu.edu}}
\textsuperscript{$\ast$} Department of Industrial and Manufacturing Engineering\\
The Pennsylvania State University, University Park, PA\\
 \textsuperscript{$\dagger$} Google Cloud, Google LLC., Mountain View, CA\\
 \textsuperscript{$\ddagger$}Department of Statistics\\ The Pennsylvania State University, University Park, PA
\end{center}
}

\maketitle

\begin{abstract}

 \citeauthor{EDCZhao_QE} (\citeyear{EDCZhao_QE}, \citeyear{zhaoEDC_Tech}, \citeyear{zhaoEDC_PE}, \citeyear{zhaoEDC_IJDS}) have recently proposed a new methodology for the Statistical Process Control (SPC) of discrete parts whose 3-dimensional (3D) geometrical data are acquired with non-contact sensors. The approach is based on monitoring the spectrum of the Laplace–Beltrami (LB) operator of each scanned part estimated using finite element methods (FEM). The spectrum of the LB operator is an intrinsic summary of the geometry of a part, independent of the ambient space. Hence, registration of scanned parts is unnecessary when comparing them. The primary goal of this case study paper is to demonstrate the practical implementation of the spectral SPC methodology through multiple examples using real scanned parts acquired with an industrial-grade laser scanner, including 3D-printed parts and commercial parts. We discuss the scanned mesh preprocessing needed in practice, including the type of remeshing found to be most beneficial for the FEM computations. For each part type, both the ``phase I" and ``phase II" stages of the spectral SPC methodology are showcased. In addition, we provide a new principled method to determine the number of eigenvalues of the LB operator to consider for efficient SPC of a given part geometry, and present an improved algorithm to automatically define a region of interest, particularly useful for large meshes. Computer codes that implement every method discussed in this paper, as well as all scanned part datasets used in the case studies, are made available and explained in the supplementary materials.
\end{abstract}


\section{Problem Description}

Consider a production process of discrete parts measured with a non-contact scanner and the corresponding problem of performing statistical process control on the acquired 3-dimensional geometrical data. The problem has received considerable attention under different simplifying assumptions, but most methods have as a common denominator the need to register, or superimpose, the parts before comparing their measurements against each other or some standard. These methods follow the tradition of Statistical Shape Analysis, in which surface measurements are aligned with the same location and orientation to facilitate statistical analysis \citep{Drydmard16}.  Once registered, the Euclidean distance between the points in each dataset is calculated to analyze significant differences between the objects \citep{EDC2010}. Despite numerous heuristics, the registration step remains computationally expensive and prone to local optima \citep{zhaoEDC_Tech},  important considerations for real-time quality control given the increasingly larger datasets acquired with non-contact sensors in the industry, with hand-held scanners commonly used. The type of data frequently provided by industrial scanners is mesh data, that is, triangulations created from the scanned clouds of points, and that will be the main focus of our discussion. 

Recently, \citeauthor{EDCZhao_QE} (\citeyear{EDCZhao_QE}, \citeyear{zhaoEDC_Tech}, \citeyear{zhaoEDC_PE}, \citeyear{zhaoEDC_IJDS}) proposed a new spectral-based methodology for the Statistical Process Control (SPC)  of 3D part geometry based on surface mesh data, which does not require part-to-part registration. The spectral SPC methods are based on an {\em intrinsic} differential geometric quantity,  i.e., they are based on attributes that depend only on the coordinates defined on the surface of each object and not on the coordinates of the scanned points measured with respect to some fixed ambient space \citep{EDCZhao_QE}.
In particular, the methodology is based on estimating the spectrum of the Laplace–Beltrami (LB) operator from the scanned triangulated meshes. The spectrum is obtained by numerically solving a Helmholtz equation, pertaining to the spatial part of the heat equation. This partial differential equation models how heat diffuses on the surface of the object that is considered. The heat equation contains the LB  differential operator, whose eigenfunctions and eigenvalues encapsulate a rich description of the geometry of the surface over which heat diffuses, in our case, the geometry of the surface of the discrete parts. In this sense, spectral SPC methods are clearly Physics-based methods. \cite{zhaoEDC_Tech,zhaoEDC_PE} suggested to use multivariate nonparametric control charts for monitoring the sequence of LB spectra of parts produced during both traditional phases in SPC: the off-line or retrospective Phase I and the online Phase II. To reduce the computational cost for very large meshes, a method to find a smaller region of interest (ROI) was further proposed using properties of the LB operator, so that the ROI is likely to contain geometrical abnormalities \citep{zhaoEDC_IJDS}. 

\begin{figure}[ht]
    \centering
    \includegraphics[scale = 0.55]{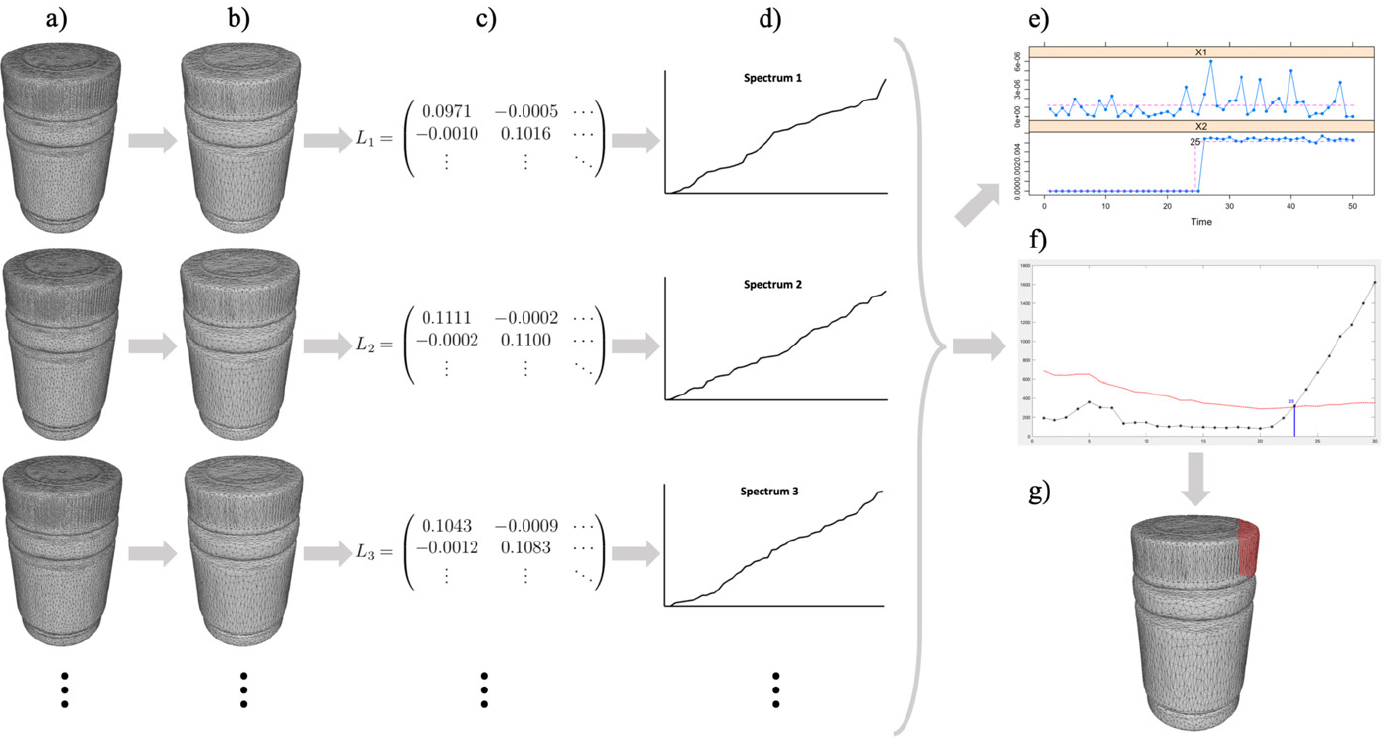}
    \caption{Spectral SPC pipeline: a) Scan objects to get meshes; b) Preprocessing of meshes; c) Estimate the LB operator for each mesh; d) Compute the lower LB spectrum; e) Phase I SPC; f) Phase II SPC; g) Post alarm diagnostics, including abnormality localization (from \cite{zhaoEDC_Tech}).}
     \label{fig: SpectralSPCPipeline}
\end{figure}

Figure \ref{fig: SpectralSPCPipeline} shows the complete ``pipeline", or different computational steps of the overall spectral SPC methodology, starting from scanned mesh data from the surface of a sequence of parts, estimation of the LB operator and its spectrum from each mesh, to Phase I and II SPC based on the LB spectrum, and finally post-alarm diagnostics. The methods were also recently extended to voxel data obtained via computed tomography (CT) scanning machines (not discussed here, we refer readers to \cite{zhaoEDC_PE}).

In this paper, we present a practical demonstration and implementation details of the spectral SPC methods with meshes obtained from a variety of real parts, either produced or purchased, scanned with an industrial-grade 3D scanner. We demonstrate the practical implementation of the entire set of techniques shown in Figure \ref{fig: SpectralSPCPipeline}, using additively manufactured parts, free-form surfaces, and commercial parts.  In addition, we provide a new principled method to determine the number of eigenvalues of the LB operator to consider for efficient SPC of a given part geometry, and present an improved algorithm based on that in \cite{zhaoEDC_IJDS} to define a region of interest in a scanned mesh, likely to contain the abnormal region in a part, useful to reduce computational cost for large meshes. The goal of the paper is therefore to help practitioners adopt spectral SPC techniques in practice.

{\bf Related work.} The spectral approach to SPC is the continuation of a line of research initiated by \cite{colosimo2008,colosimo2014profile} (see also \cite{colosimo2018}) on statistical process control methods for geometric features. As these authors indicate, ``When the quality of a manufactured product is related to geometric specifications (e.g., profile and form tolerances as straightness, roundness, cylindricity, flatness, etc.), the process should be considered in-control if the relationship used to represent that profile or surface in the space is stable over time” \citep{colosimo2008}. Rather than monitoring individual geometrical features of a part, of which a complex part will have many, spectral SPC methods help to monitor the {\em complete} geometry of the part, which the spectrum of the Laplace-Beltrami operator encapsulates. This permits, in principle, the monitoring of the stability over time of all the geometrical features from a sequence of parts with a single multivariate SPC chart. More importantly, while there have been previous approaches to monitor whole surface data, almost all the previous literature on SPC for geometrical features is based on registrating the parts either between them or against a CAD model (e.g., see \cite{wang2014} and for a recent review of these approaches, which are based on a variety of statistical techniques once parts are registered, see \cite{zhao2022}).
 As mentioned before, a main motivation of the spectral SPC methods is to monitor intrinsic geometrical features of the parts, that is, features that do not depend on the orientation or location of the parts in 3D Euclidean space and hence, completely avoid the registration step of earlier approaches.\\

The rest of the paper is organized as follows. Section 2 presents practical aspects of the data collection and preparation, particularly the mesh preprocessing steps needed to implement the spectral methods from data acquired using industrial-grade scanners. Next, in Section \ref{sec: No. Of Eig-vals}, we discuss the problem of how to choose the number of eigenvalues. Section \ref{sec: SPC methods} presents the implementation of the SPC charting methods (for both ``phase I" and ``phase II") that are based on the estimated LB spectrum. Then, Section \ref{sec: ROI} presents a new method for finding an ROI on a given scanned mesh to reduce computational cost. The resulting ROI is selected such that it is likely to contain abnormalities. The method is illustrated with real scanned parts and is useful when it is difficult for an operator or engineer to define an ROI based on process experience. Section \ref{sec: summary} ends the paper with a summary of advice to practitioners. All computer codes that implement the various spectral SPC methods and the mesh datasets used in the case study are made available as supplementary materials.

\section{Data Collection and Preparation}
\label{sec: pre-processing}

Preprocessing of the meshes acquired from non-contact laser scanners is always necessary to clean the meshes of extraneous artifacts due to the scanning procedure, such as outlier noise and even the surface on which the part is located, which will be scanned as well. In addition, experience with Finite Element Methods (FEM) shows how isotropic remeshing algorithms (IRA) \citep{alliez2003isotropic} help improve the numerical stability of FEM computations.  

\setcounter{figure}{1}
\begin{figure}[ht]
    \centering
    \begin{subfigure}{0.45\textwidth}
    \centering
    \includegraphics[height = 4cm, width = 6cm]{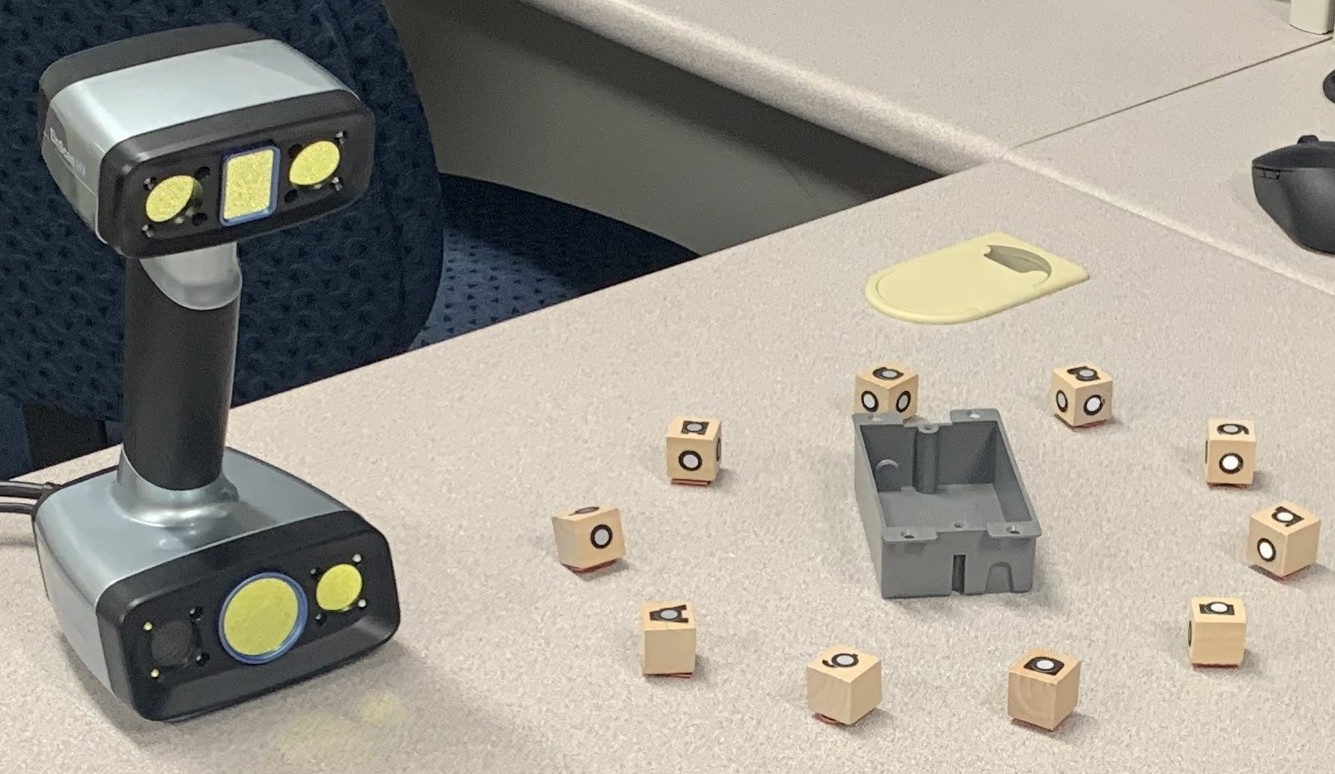}
  \end{subfigure}
  \hspace{0.5cm}
  \begin{subfigure}{0.45\textwidth}
  \centering
    \includegraphics[height = 4cm, width = 5cm]{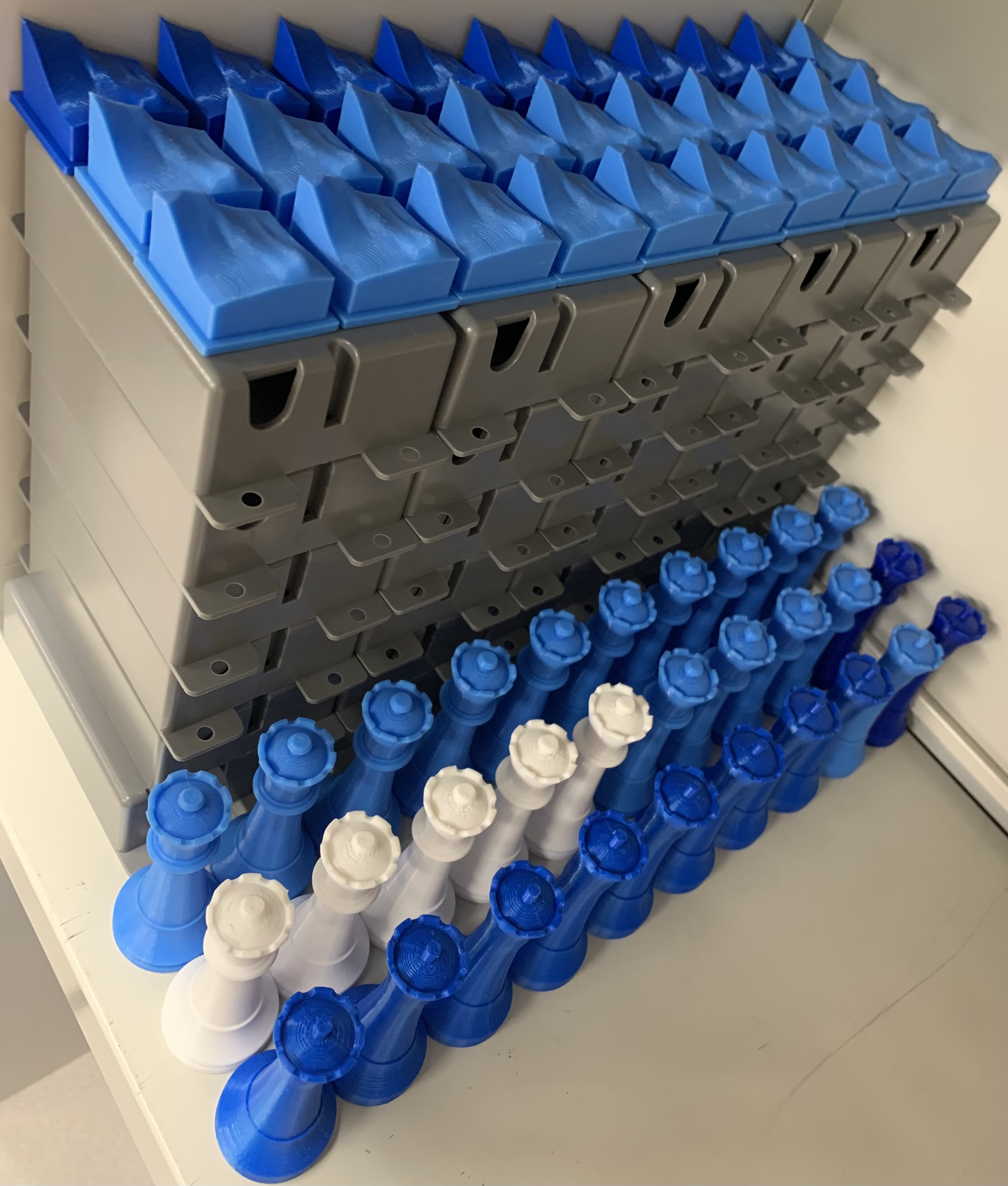}
  \end{subfigure}
  \caption{On the left, the experimental setup, showing one of the parts scanned, and on the right, the set of all parts scanned during our experiments.}\vspace{0.5cm}
    \label{fig: Set_up}
\end{figure}

\begin{figure}[ht]
    \centering
    \includegraphics[scale = 0.5]{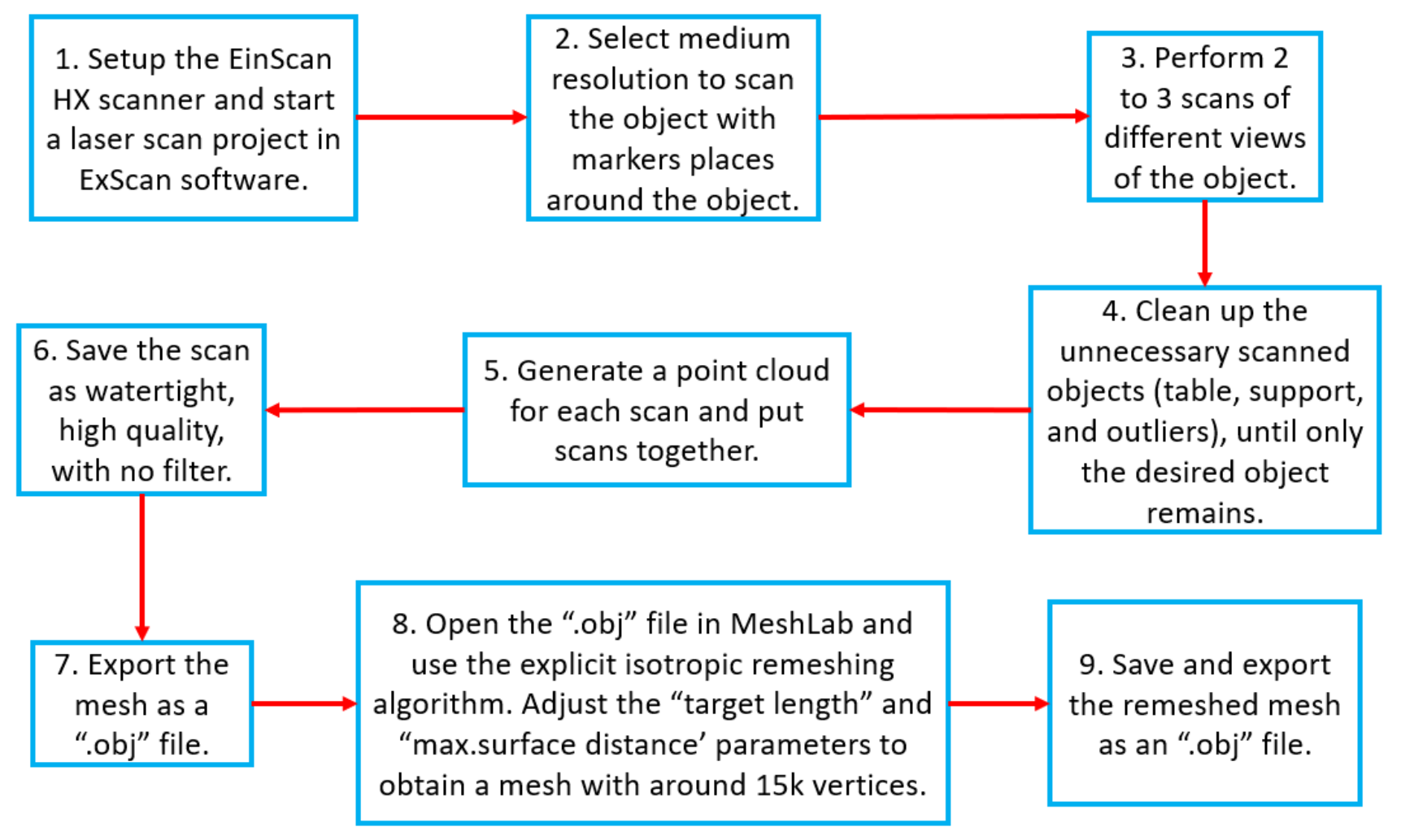}
    \caption{Mesh acquisition and preprocessing procedure, until the creation of the resulting mesh file. These steps were applied to all parts scanned for this paper. }
    \label{fig: PP_flowchart}
\end{figure}

Figure \ref{fig: Set_up}  shows the experimental setup of the scanning procedure performed on all the parts used in this paper, which are shown on the right of the figure. We acquired and preprocessed each mesh following the procedure indicated in Figure \ref{fig: PP_flowchart}. Note how this includes IRA preprocessing. A SHINING3D EinScan HX industrial-grade hand-held scanner was used together with the EXScan software interface. However, steps 2 to 7 in figure \ref{fig: PP_flowchart} can be implemented with scanning software accompanying any other industrial or commercial scanner. We also used the open-source MeshLab software \citep{Meshlab} to perform the IRA remeshing.  To enhance the accuracy of the scanning process, small cubes with markers on three of their faces were placed around the objects to be scanned, as shown on the left of Figure \ref{fig: Set_up}.   A thorough discussion and justification regarding IRA is provided in Appendix \ref{app 1}. 

\section{Analysis and Interpretation}
\subsection {Choosing the Number of Eigenvalues}
\label{sec: No. Of Eig-vals}
An important practical problem left unanswered in the literature of spectral SPC methods is how to determine how many of the $N$ leading eigenvalues (where $N$ is the number of vertices in a mesh) to include in the LB spectrum. The method we propose for doing this is based on an analogy with Fourier analysis. Similarly to the Fourier analysis of a function which is approximated by different harmonics of different frequencies, with low frequency harmonics describing the general shape of a function and higher frequencies adding more details, the leading eigenfunctions of the LB operator describe the general shape of a surface, as they correspond to eigenvalues representing low frequencies of vibration of a 3-dimensional ``membrane", while the higher eigenfunctions and eigenvalues relate to high-frequency membrane vibrations which capture minor additional geometrical details of the surface geometry. The surface is seen as the result of the superposition of vibrating membranes at ``pure tones", as in the Fourier analysis of a function (given that we work with an estimated LB operator from a mesh representation of a surface, rather than a continuous surface as in mathematics, this results in eigen{\em vectors} rather then eigen{\em functions}). 

Given a part mesh $\mc{M}$ with $N$ vertices, it is possible to perfectly reconstruct it by considering {\em all} $N$ eigenvectors of its estimated LB operator (so, in our methods, the monitored LB spectrum would include all $N$ eigenvalues). This is clearly impractical, as meshes acquired with industrial-grade scanners can have hundreds of thousands of points. However, just as in Fourier approximations of a function, which are truncated after a certain number of harmonics, we could do an approximate reconstruction of $\mc{M}$   using only the first $k < N$ eigenvectors, and determine how different the reconstructed mesh is from the noise-free Computer-Aided-Design (CAD) model (assumed given in the form of a mesh) for increasing values of $k$.  To determine the best value of $k$, we can compute the Frobenius norm distance of the point-to-point differences between the meshes (CAD vs. reconstructed), and plot these distances against $k$, in a plot analogous to the ``scree" plot popular in classical principal component analysis \citep{cattell1966scree}.

The details of the method used to reconstruct the CAD model from its first $k$ eigenvectors are given in Appendix \ref{app_Reconstruct}.

For a given number of eigenvectors $k$, with $1 <k \leq N$, define the matrix coordinate distance between the reconstructed mesh, $\mc{M}_k$, and the CAD model mesh $\mc{M}$, as $D_k = P_k-P_0=[d_{ij}]$.  We then compute the Frobenius norm of this matrix, i.e., 
\[d(\mathcal{M}_k, \mathcal{M}_0) = \| D_k \|_F = \sqrt{\sum_{i=1}^{N}\sum_{j=1}^{3} d_{ij}^2}\]

The values of $d(\mathcal{M}_k,\mathcal{M}_0)$  can be plotted against the values of $k$ to give a ``scree"-like plot. The selection of the value $k$ is based on finding the ``elbow" in this plot, similarly to principal component analysis. 

Figures \ref{fig: Queen Chess Piece Scree Plot} to \ref{fig: Freeform Scree Plot} demonstrate the proposed method for three different types of parts: a 3D-printed chess piece, a commercial electrical box, and a 3D-printed free-form surface. In each figure, the left panels display the reconstructed meshes using an increasing number $k$ of LB eigenvectors. As discussed in \cite{zhaoEDC_IJDS}, to reconstruct objects with a relatively simple shape, it is sufficient to use the first few tens of eigenvectors. This can be further confirmed in the right panels where $\|D_k\|_F$ is plotted against $k$. One can observe the steep drop in the Frobenius norm distance when using the first leading eigenvectors, after which  $\|D_k\|_F$ flattens out for larger $k$. This indicates that implementing the spectral SPC methods with just the first 15 to 30 leading eigenvalues is enough for monitoring these three types of parts.
\setcounter{figure}{3}
\begin{figure}[ht]
    \centering
    \begin{subfigure}{0.45\textwidth}
    \centering
    \includegraphics[scale = 0.45]{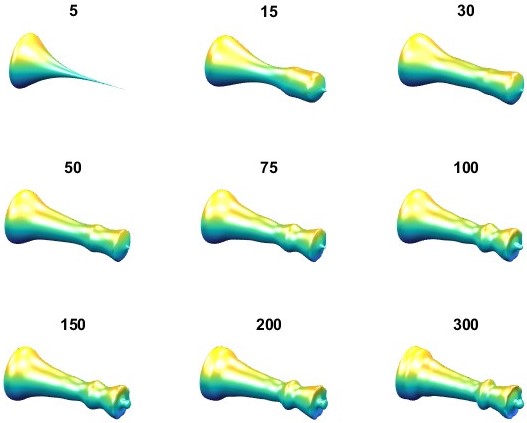}
  \end{subfigure}
  \hspace{0.5cm}
  \begin{subfigure}{0.45\textwidth}
  \centering
    \includegraphics[scale = 0.375]{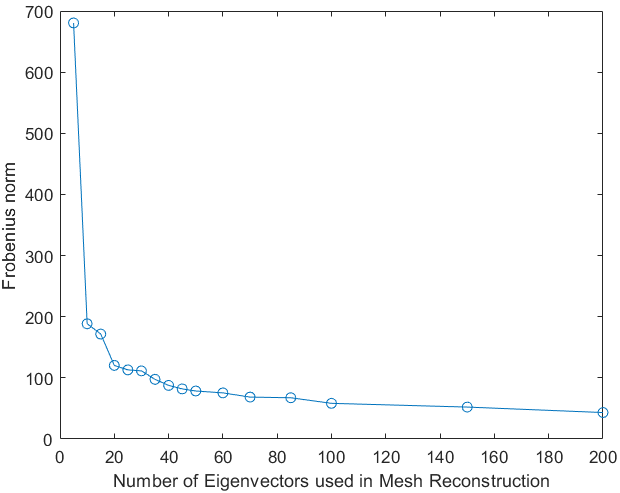}
  \end{subfigure}
    \caption{Left: Reconstructed CAD mesh of a 3D-printed chess piece (a queen) with different numbers of eigenvectors. Right: The plot of the Frobenius norm of the difference matrix, $\|D_k\|_F$, against the number of eigenvectors in the reconstruction, $k$.  }
    \label{fig: Queen Chess Piece Scree Plot}
\end{figure}
\setcounter{figure}{4}
\begin{figure}[ht]
    \centering
    \begin{subfigure}{0.45\textwidth}
    \centering
    \includegraphics[scale = 0.45]{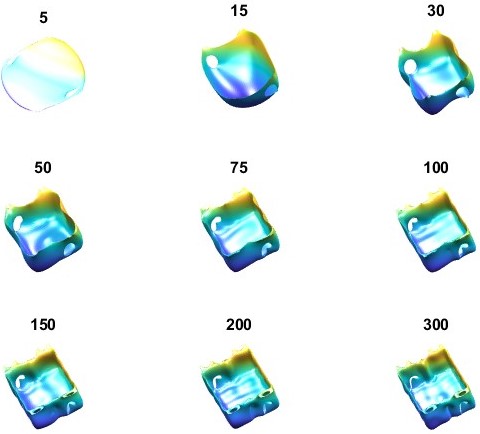}
  \end{subfigure}
  \hspace{0.25cm}
  \begin{subfigure}{0.45\textwidth}
  \centering
    \includegraphics[scale = 0.375]{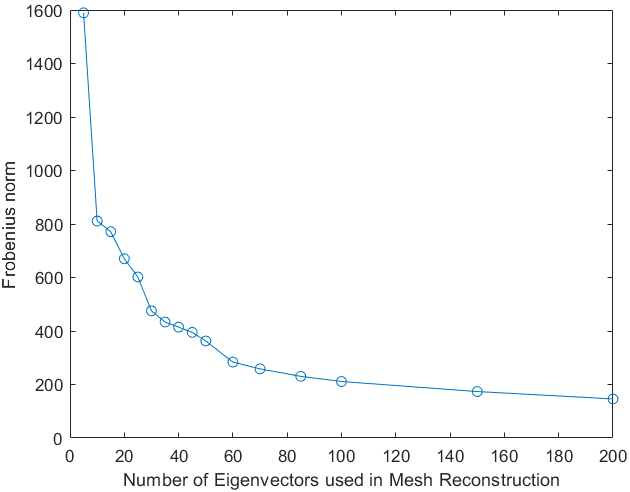}
  \end{subfigure}
    \caption{Left: Reconstructed CAD mesh of a commercial electrical box with different numbers of eigenvectors. Right: The plot of the Frobenius norm of the difference matrix, $\|D_k\|_F$, against the number of eigenvectors in the reconstruction, $k$. }
    \label{fig: Elec Box Scree Plot}
\end{figure}
\setcounter{figure}{5}
\begin{figure}[ht]
    \centering
    \begin{subfigure}{0.45\textwidth}
    \centering
    \includegraphics[scale = 0.45]{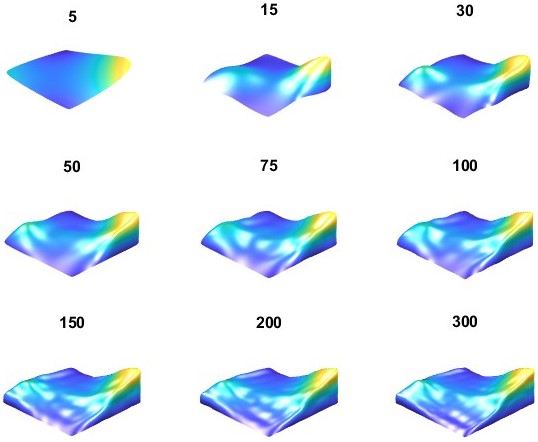}
  \end{subfigure}
  \hspace{0.5cm}
  \begin{subfigure}{0.45\textwidth}
  \centering
    \includegraphics[scale = 0.375]{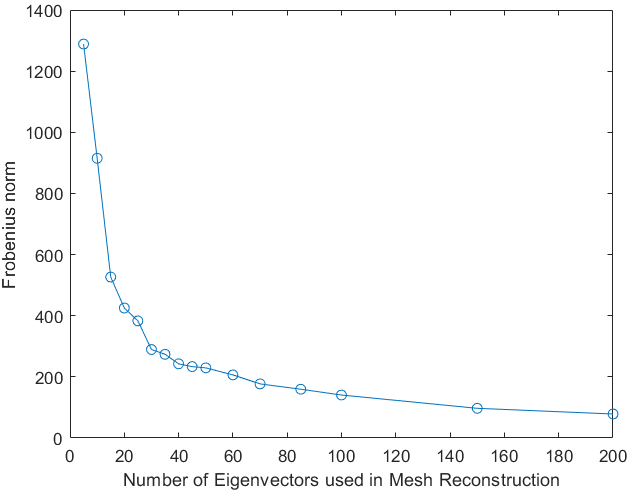}
  \end{subfigure}
    \caption{Left: The reconstructed CAD mesh of a free-form surface with different numbers of eigenvectors. Right: The plot of the Frobenius norm of the difference matrix, $\|D_k\|_F$, against the number of eigenvectors in the reconstruction, $k$. }
    \label{fig: Freeform Scree Plot}
\end{figure}

\subsection{Demonstration of Distribution-free Multivariate SPC Methods Based on the LB Operator Spectrum}
\label{sec: SPC methods}

\cite{zhaoEDC_Tech} showed via simulation that even if the noise added to a surface was normally distributed, the estimated LB operator spectrum (a vector) is not multivariate normal. Hence, given the non-normal nature of the LB spectrum, they suggested using distribution-free multivariate control schemes for both phase I (retrospective testing) and phase II (online monitoring) in their spectral SPC framework, and applied them to the lower part of the estimated LB spectra. For phase I, a multivariate permutation-based method based on signed ranks-related statistics proposed by \cite{capizzi2017phase}, available in their R package \texttt{dfphase1} (function $\tt mphase1$) is used. For phase II, \cite{zhaoEDC_Tech} adapted a distribution-free multivariate EWMA control chart, proposed by \cite{chen2016distribution}, which is based on permutation tests on Wilcoxon rank-sum statistics (\cite{zhaoEDC_Tech} provide the corrected moments of the monitored statistic in their paper).  For a detailed discussion on the detection performance of these phase I and phase II methods applied to the spectrum of the LB operator, and comparisons with alternative registration-based SPC methods, see \cite{zhaoEDC_Tech,zhaoEDC_PE}. Here we confine ourselves to demonstrating the practical aspects of the two phases using sequences of real scanned (and preprocessed) parts.

\subsubsection{Demonstration of phase I SPC using the LB operator spectrum (retrospective testing)}

We demonstrate the phase I spectral SPC methods with the three real parts shown before:  3D-printed queen chess pieces, commercial electrical boxes (purchased), and 3D-printed free-form surfaces. 
In this demonstration and that of the next section, we formed sequences of 3D-printed (or purchased) parts, produced according to their nominal CAD model geometry (herein ``{\em \bf normal parts}"). In addition, to create an ``out-of-control" state of the process, we introduced subsequences of parts with small changes in their geometry (with respect to their CAD model) which we will refer herein as ``{\em \bf abnormal parts}" (note that according to common practice, the spectral SPC methods rely only on part measurements and not on CAD specifications). The changes in the geometry related to abnormal parts are highlighted with red circles in the left panels of Figures \ref{fig: Phase II SPC Queen Chess Piece} to \ref{fig: Phase II SPC Freeform}. These reproduce manufacturing errors frequently found in 3D printing (chess pieces and free-form surfaces) or in injection-molded plastic manufacturing (electrical boxes). In all SPC demonstrations (for both phase I and phase II), we first preprocessed the scanned meshes as discussed above, down-sampling to approximately $N$=15K vertices per mesh.

For retrospective testing (phase I), let $m_0$ be the number of parts collected. We used $m_0 = 20$ in all three examples. We initially considered the case where all 20 parts were produced while the process was in-control, producing only normal parts. All parts were scanned and preprocessed as described in Section  \ref{sec: pre-processing}, and for each scanned mesh, the first $k$ LB eigenvalues were computed via FEM \citep{zhaoEDC_PE} where $k$ was selected using the reconstruction method presented in the previous section (in this demonstration, $k=15$ was selected for all three part types). The time-ordered sequence of estimated LB spectra vectors for each part type was then entered into function $\tt mphase1$ (all default parameters of this function were used), and, accordingly, no out-of-control state was detected during the 20-run-long phase I for any of the part types.  We next tried the case where, for each of the three types of parts, the first 10 parts (parts 1 to 10) were produced under an in-control state while the last 10 parts (parts 11 to 20) were produced after a sustained or step shift to an out-of-control state resulting in abnormal parts. In this case, the alarm probabilities returned by $\tt mphase1$ are as reported in Table \ref{tab: Phase I}. 

Table \ref{tab: Phase I} indicates the eigenvalue number(s) deemed out-of-control during phase I and when the sustained shift was estimated to occur. In the first two cases, for the chess pieces and the free-from surfaces, the shift time is precisely estimated to occur at time 10. For the electrical boxes, it is estimated to be at time 9 (as opposed to time 10). As can be seen, the overall phase I procedure works remarkably well for all 3 part types in this demonstration. Evidently, a full phase I detection performance analysis of the method cannot be done with real parts (and the same can be said about a phase II run length analysis). The performance of the spectral SPC methods depends on the size of the geometric abnormalities relative to the mesh resolution, the size of the part, and the level of noise present.   Details of the excellent phase I performance of the \cite{capizzi2017phase} approach when applied to the LB spectrum of (simulated) surfaces are given in \cite{zhaoEDC_Tech}.

\begin{table}[htbp]
    \centering
    \caption{Phase I SPC results for the three types of parts, listing the eigenvalues deemed out-of-control, the estimated point in time at which the out-of-control state was detected, and the reported p-values of the detection test.}
    \begin{tabular}{| c|c|c|c|}
        \hline
        & \textbf{Chess piece} & \textbf{Free-form surface} & \textbf{Electrical box}\\
        \hline
        Detected eigs. & Time 10: $X_5$ & Time 10: $X_1, X_3,X_5, X_9$ & Time 9: $X_3,X_8,X_{12}$ \\
        \hline
        p-value & 0.009 & 0.009 & 0.01 \\
        \hline
    \end{tabular}
    \label{tab: Phase I}
\end{table}

\subsubsection{Demonstration of phase II SPC using the LB operator spectrum (on-line monitoring)}

We next use the same three real part types shown in the previous section to demonstrate online monitoring (phase II).   For each of the three examples, a sequence of 30 parts was produced (or purchased) and scanned. These included 20 normal parts and an additional 10 abnormal parts. All parts were scanned and processed according to the procedure described in Section \ref{sec: pre-processing}. Figures \ref{fig: Phase II SPC Queen Chess Piece} to \ref{fig: Phase II SPC Freeform}  show instances of a normal and an abnormal part with the abnormality circled in red. Figures \ref{fig: Phase II SPC Queen Chess Piece} to \ref{fig: Phase II SPC Electrical box}  also display the corresponding phase II control chart (this is the DFEWMA chart in \cite{chen2016distribution} using $m_0=20$ and the corrected first and second moments of the test statistic as given by \cite{zhaoEDC_Tech}).

A sustained out-of-control state was introduced at the beginning of the online monitoring phase ($t \geq 1$), assuming it results in the types of abnormalities displayed in figures \ref{fig: Phase II SPC Queen Chess Piece}-\ref{fig: Phase II SPC Freeform}. As can be seen from these figures, for the 3D-printed queen chess pieces, 3D-printed free-form surfaces, and commercial electrical boxes examples, the phase II method rapidly detected the process shift at $t=2$, $t=4$, and $t=4$ respectively.

\setcounter{figure}{6}
\begin{figure}[ht]
    \centering
    \begin{subfigure}{0.4\textwidth}
    \centering
    \includegraphics[scale = 0.5]{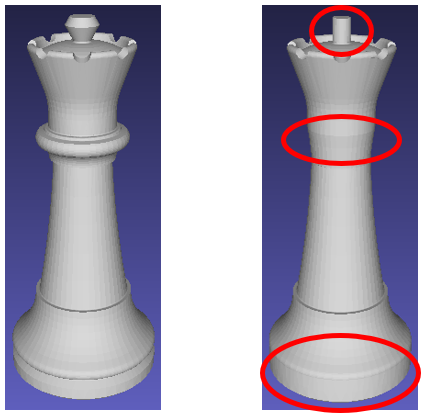}
  \end{subfigure}
  \begin{subfigure}{0.4\textwidth}
  \centering
    \includegraphics[scale = 0.3]{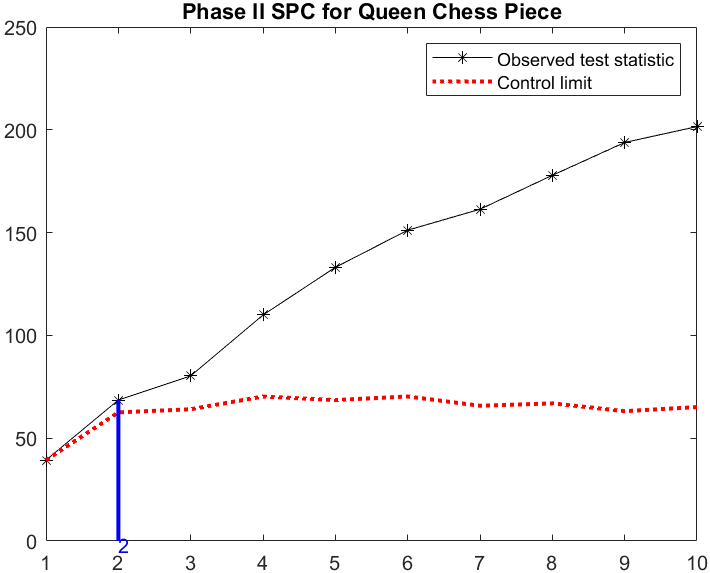}
  \end{subfigure}
    \caption{Left: In-control and out-of-control produced queen chess pieces, the latter with abnormalities encircled in red. Right: a permanent shift to an out-of-control state was introduced starting at $t=1$, and the phase II control chart by \cite{chen2016distribution} (with the corrected moments for the test statistics as given in \cite{zhaoEDC_Tech}) detected the resulting abnormal parts immediately after, at $t=2$.}
    \label{fig: Phase II SPC Queen Chess Piece}
\end{figure}

\setcounter{figure}{7}
\begin{figure}[ht]
    \centering
    \begin{subfigure}{0.4\textwidth}
    \centering
    \includegraphics[scale = 0.325]{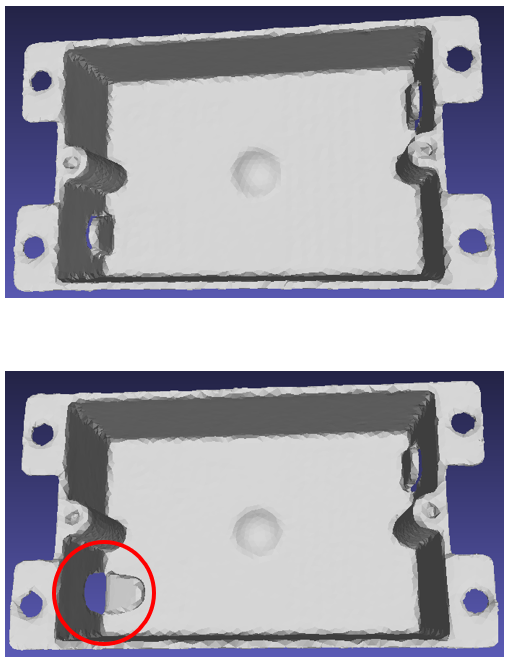}
  \end{subfigure}
  \begin{subfigure}{0.4\textwidth}
  \centering
    \includegraphics[scale = 0.45]{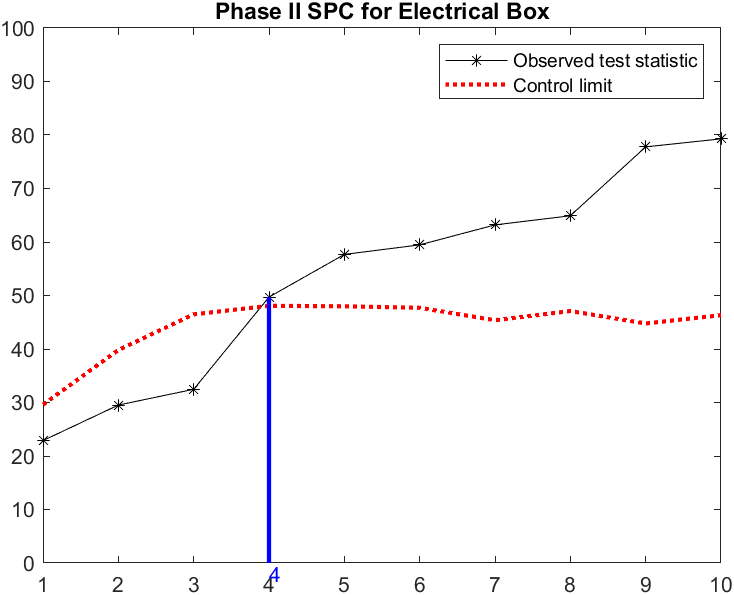}
  \end{subfigure}
    \caption{Left: In-control and out-of-control electrical boxes with an abnormality encircled in red. Right: a permanent shift to an out-of-control state was introduced starting at $t=1$, and the phase II control chart by \cite{chen2016distribution} (with the corrected moments for the test statistics as given in \cite{zhaoEDC_Tech}) detected the resulting abnormal parts at $t=4$.}
    \label{fig: Phase II SPC Electrical box}
\end{figure}
\setcounter{figure}{8}
\begin{figure}[ht]
    \centering
    \begin{subfigure}{0.4\textwidth}
    \centering
    \includegraphics[height = 4.5cm, width = 3cm]{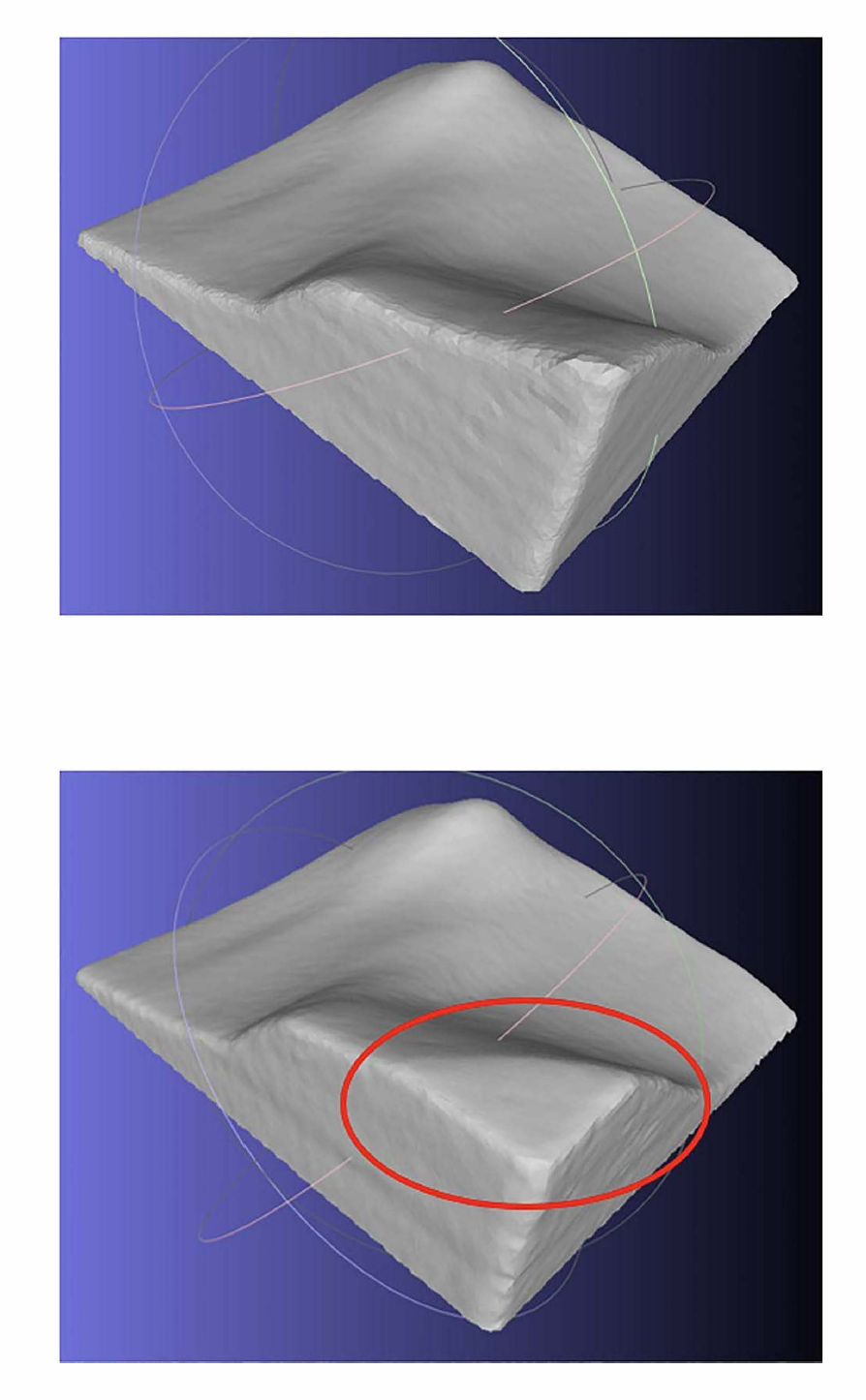}
  \end{subfigure}
  \begin{subfigure}{0.4\textwidth}
  \centering
    \includegraphics[scale = 0.35]{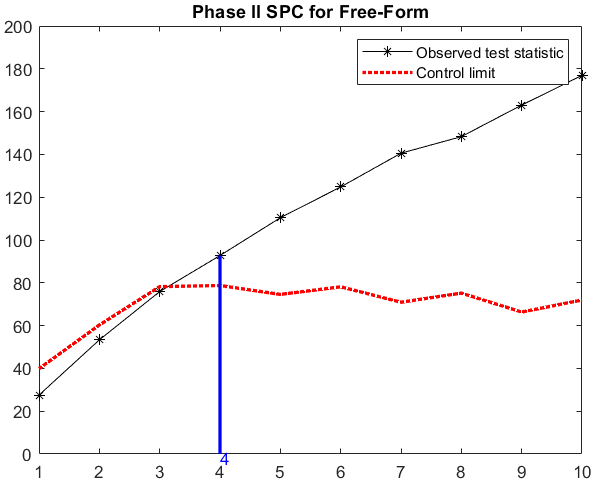}
  \end{subfigure}
    \caption{Left: In-control and out-of-control produced free-form surfaces, the latter with an abnormality encircled in red. Right: a permanent shift to an out-of-control state was introduced starting at $t=1$, and the phase II control chart by \cite{chen2016distribution} (with the corrected moments for the test statistics as given in \cite{zhaoEDC_Tech}) detected the resulting abnormal parts at $t=4$.}
    \label{fig: Phase II SPC Freeform}
\end{figure}

Readers interested in a formal average run length performance of the method should consult \cite{zhaoEDC_Tech}, who showed how the spectral SPC method is effective in detecting small changes in the shape or size of parts.

\subsubsection{Demonstration of a post-alarm diagnostic: abnormality localization on a given surface}

 In practice, once an alarm is triggered during phase II, assuming it is not a false alarm,  it is not known what type of geometrical abnormalities caused it, since the LB spectrum considers the whole geometry of a part, and whether this is severe enough for the parts leading to the alarm to be considered defective from an {\em inspection} point of view. In this section, we assume a user has followed the \cite{chen2016distribution} method to determine the part where the changepoint in the process occurred. Hence, the next task is to locate the abnormality on the meshes corresponding to the parts after where the changepoint occurred, to help an investigation of the underlying assignable causes that would lead to the improvement of the process, following the usual Shewhart SPC paradigm. To perform this task, \cite{zhaoEDC_Tech} proposed a method based on the well-known iterative closest point (ICP) algorithm \citep{besl1992method}, used to simply register the mesh of the (noise-free) CAD model of the part, with the mesh corresponding to the suspected part. As previously mentioned in the introduction, registration methods depend on a good initial alignment, otherwise, the optimization algorithm will easily get stuck at a local minimum, which results in the failure of the ICP registration. Given that there is no guarantee for a good initial alignment, especially if portable scanners are used, the ICP registration could be unreliable for the post-alarm diagnostics. Therefore, when the ICP algorithm failed to register part and CAD model successfully, the $\tt{Align}$ function in MeshLab \citep{Meshlab} was employed to manually align a few points on the abnormal part that triggered the alarm with the corresponding points on the CAD model to facilitate accurate ICP registration. This is essentially the same procedure followed by commercial inspection software that relies on ICP registration. Note ICP registration is only used in a diagnostic step once detection has occurred, and not for SPC (we note how recently, \cite{zhaoEDC_IJDS} present an alternative diagnostic method for the localization of an abnormality on the surface of a part that does {\em not} rely on registration of part and CAD meshes -- it uses the concept of a functional map -- making this diagnostic also registration-free).

 Figure \ref{fig: Diagnostics} shows the location of the abnormalities on all three example parts used in the previous sections by the usual device of coloring the surface of the noise-free CAD model according to the deviation from the nominal models, with light yellow color corresponding to the largest deviations. Upon successful registration, abnormalities in all three parts are correctly and accurately identified. 
\setcounter{figure}{9}
\begin{figure}[ht]
    \centering
    \begin{subfigure}{0.3\textwidth}
    \centering
    \includegraphics[scale = 0.25]{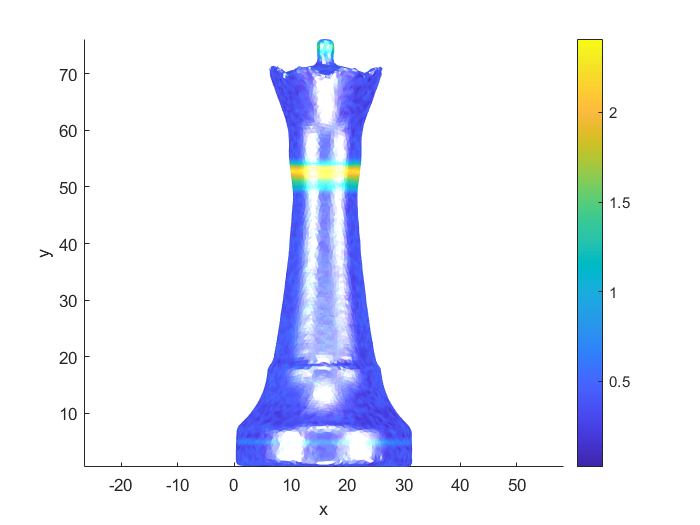}
  \end{subfigure}
  \begin{subfigure}{0.3\textwidth}
  \centering
    \includegraphics[scale = 0.25]{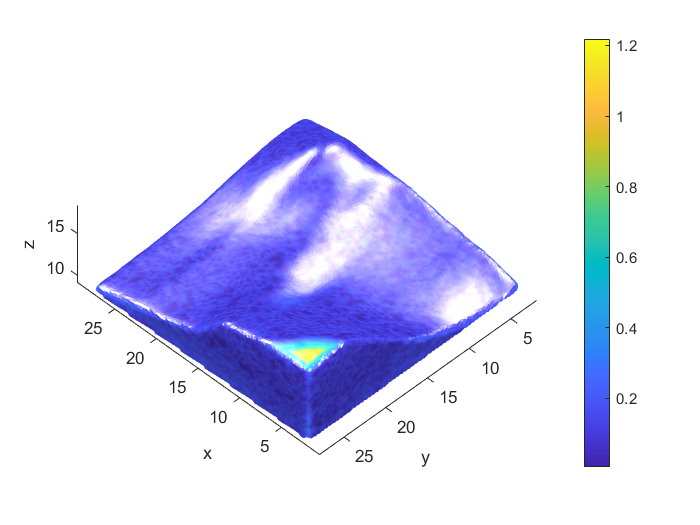}
  \end{subfigure}
  \begin{subfigure}{0.3\textwidth}
  \centering
    \includegraphics[scale = 0.25]{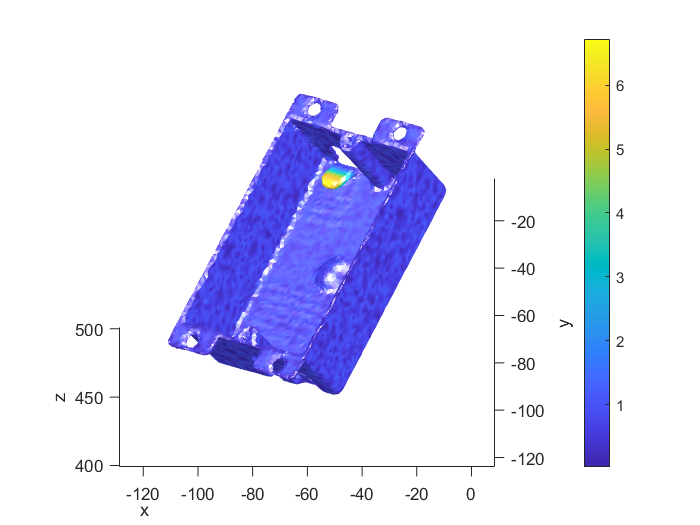}
  \end{subfigure}
    \caption{Post-alarm diagnostic to localize abnormalities on the surface of a part for three example parts. Lighter areas indicate greater deviations from the CAD model.}
    \label{fig: Diagnostics}
\end{figure}

Practitioners interested in the run length performance of the DFEWMA chart applied to the LB spectrum, considering different types and sizes of various geometric abnormalities, under different levels and types of noise, should consult \cite{zhaoEDC_Tech}. 

\subsection{Demonstration of a New Algorithm for Defining a Region of Interest}
\label{sec: ROI}
 Recently, \cite{zhaoEDC_IJDS} developed a recursive algorithm for subdividing a part surface and finding a smaller Region of Interest (ROI)  where abnormalities are likely to occur, in case such an ROI cannot be defined by engineers from past process experience.  The method is based on finding a functional map between the CAD model (in the form of a mesh) and the real part scanned mesh.  The method can be used to reduce the computational demands of all spectral SPC tasks presented in this paper. Evidently, working with smaller meshes has the advantage of significantly reducing the computational cost associated with solving the required generalized eigenvalue problem. Here we demonstrate the technique using real scanned meshes, pointing out practical difficulties when implementing this method in practice, and how to solve them.

The main idea of the algorithm by \cite{zhaoEDC_IJDS} is to recursively partition the mesh of a part suspected to have an abnormal region (mesh $\mathcal{A}$) into two connected components (using their {\em nodal domains}, see Appendix \ref{app 2}) and select the component that deviates the most from its {\em corresponding} component in the noise-free CAD model (mesh $\mathcal{B}$). A binary search tree is then formed for both the CAD model and the actual scanned mesh, partitioning at each iteration the mesh of the selected component into two submeshes. The small discrepancies in the original two meshes, for the CAD model and the scanned part mesh, tend to accumulate over several iterations down the binary tree, and therefore the algorithm should be applied recursively for at most three iterations, or earlier if a small enough ROI has been obtained, with the last subcomponent deviating the most from the corresponding one in the CAD model selected to define the ROI.

Denote by $\mathcal{A}_+$ (or $\mathcal{B}_+$) and $\mathcal{A}_-$ (or $\mathcal{B}_-$)  the nodal domains defined by the positive and negative values for the second LB eigenvector of mesh $\mathcal{A}$ (or $\mathcal{B}$, respectively). Due to the ambiguity of the sign of the eigenvectors, $\mathcal{A}_+$ (or $\mathcal{A}_-$) does not necessarily correspond to the same region as $\mathcal{B}_+$ (or $\mathcal{B}_-$). Hence, to correctly match the components of $\mathcal{A}$ with the components of $\mathcal{B}$, \cite{zhaoEDC_IJDS} simply compared the mesh size of each component to determine the correspondences between the two partitions of $\mathcal A$ and the two partitions of $\mathcal B$.  

However, in practice, due to the geometry of the abnormalities and the nature of the scanning operation, similar meshes acquired with a non-contact laser scanner after applying IRA do not always have a similar distribution of their vertices on the mesh surface. In other words, it is insufficient to simply compare the mesh size of each component because of the uniquely distributed vertices on each mesh. Therefore, we propose the following modifications to the ROI algorithm presented in \cite{zhaoEDC_IJDS} to find the corresponding submeshes in both a manufactured part and the CAD model where abnormalities (with respect to the CAD model) are located.

First, compute the following differences between the {\em scaled} eigenvalues of all four pairs of submeshes:
\begin{align*}
    & d_1 := \sum_{i=2}^{15} \left| \frac{\lambda_{+, i}^A}{\mu_+^A} - \frac{\lambda_{+, i}^B}{\mu_+^B} \right|, \quad d_2 := \sum_{i=2}^{15} \left| \frac{\lambda_{+, i}^A}{\mu_+^A} - \frac{\lambda_{-, i}^B}{\mu_-^B} \right|, \\
    &  d_3 := \sum_{i=2}^{15} \left| \frac{\lambda_{-, i}^A}{\mu_-^A} - \frac{\lambda_{+, i}^B}{\mu_+^B} \right|, \quad d_4 := \sum_{i=2}^{15} \left| \frac{\lambda_{-, i}^A}{\mu_-^A} - \frac{\lambda_{-, i}^B}{\mu_-^B} \right|,
\end{align*}
where $\lambda^A_{+, i}$ denotes the $i^{th}$ LB eigenvalue of submesh $\mathcal{A}+$ and
\[\mu_+^A = \left( \prod_{j=2}^{15} \lambda_{+,j}^A \right)^{\frac{1}{14}}\]
represents their geometric mean. Similar notations apply to eigenvalues and geometric means of other submeshes. Finally, given the pair of submeshes that result in the smallest distance, we choose their corresponding submesh {\em complements} as the pair of submeshes to consider in each iteration of the algorithm since they are therefore the most dissimilar {\em corresponding} submeshes, hence likely to contain an abnormality (with respect to the CAD model). Comparing the spectrum of two surfaces needs to be done in a scaled manner because it is known that the asymptotic slope of the LB spectrum vs. eigenvalue index graph is proportional to the surface area \citep{zhaoEDC_Tech}. Thus, comparing non-scaled eigenvalues can lead the algorithm to find the wrong correspondences between meshes. Ideally, one would therefore use the surface area for normalizing the spectrum of each submesh, but this would be very computationally expensive. Instead, we can scale the first $k$  eigenvalues by their arithmetic or geometric mean, with the latter preferred because $\lambda_i>0$ for index $i$ greater equal to 2. In the demonstration of the method shown, $k$=15 was used for all three part types, where the value of $k$ is determined with the method shown in section 3.1.

\begin{figure}[ht]
    \centering
    \includegraphics[scale = 0.4]{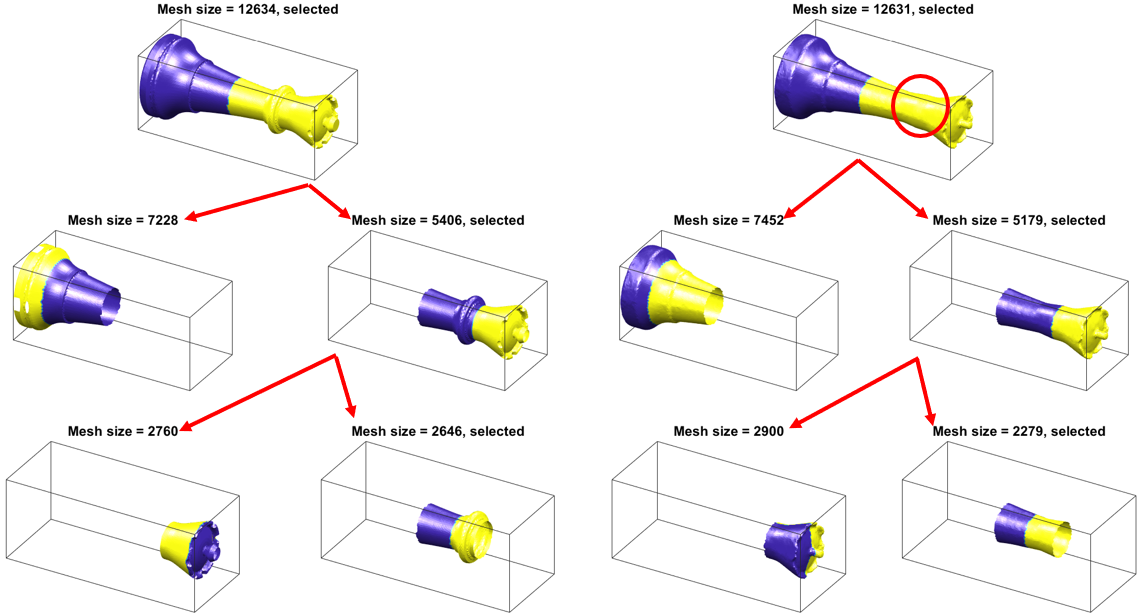}
    \caption{Recursive partitioning applied to a 3D-printed queen chess piece to define a region of interest. Each row corresponds to an iteration, where starting from row 2, columns 1 to 4 represent sub-meshes $B+$, $B-$, $A+$, and $A-$, respectively. The red arrows indicate how the selected pair of components is further partitioned in the next iteration. Colors indicate the sign of the second LB eigenvector (yellow for positive and purple for negative). The algorithm correctly selects in the last iteration the submesh that contains the true local abnormality, circled in red on the top row, which is the ROI found.}
    \label{fig: ROI of Queen Chess Piece}
\end{figure}

\begin{figure}[ht]
    \centering
    \includegraphics[scale = 0.45]{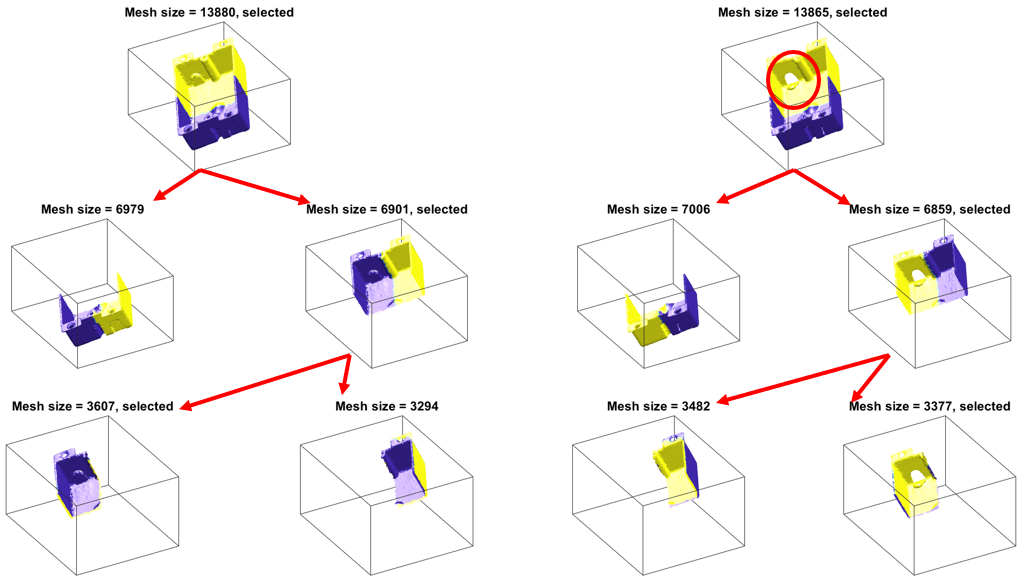}
    \caption{Recursive partitioning applied to a commercial electrical box part to define a region of interest. Each row corresponds to an iteration, where starting from row 2, columns 1 to 4 represent sub-meshes $B+$, $B-$, $A+$, and $A-$, respectively. The red arrows indicate how the selected pair of components is further partitioned in the next iteration. Colors indicate the sign of the second LB eigenvector (yellow for positive and purple for negative). The algorithm correctly selects in the last iteration the submesh that contains the true local abnormality, circled in red on the top row, which is the ROI found.}
    \label{fig: ROI of Elec Box}
\end{figure}

\begin{figure}[ht]
    \centering
    \includegraphics[scale = 0.4]{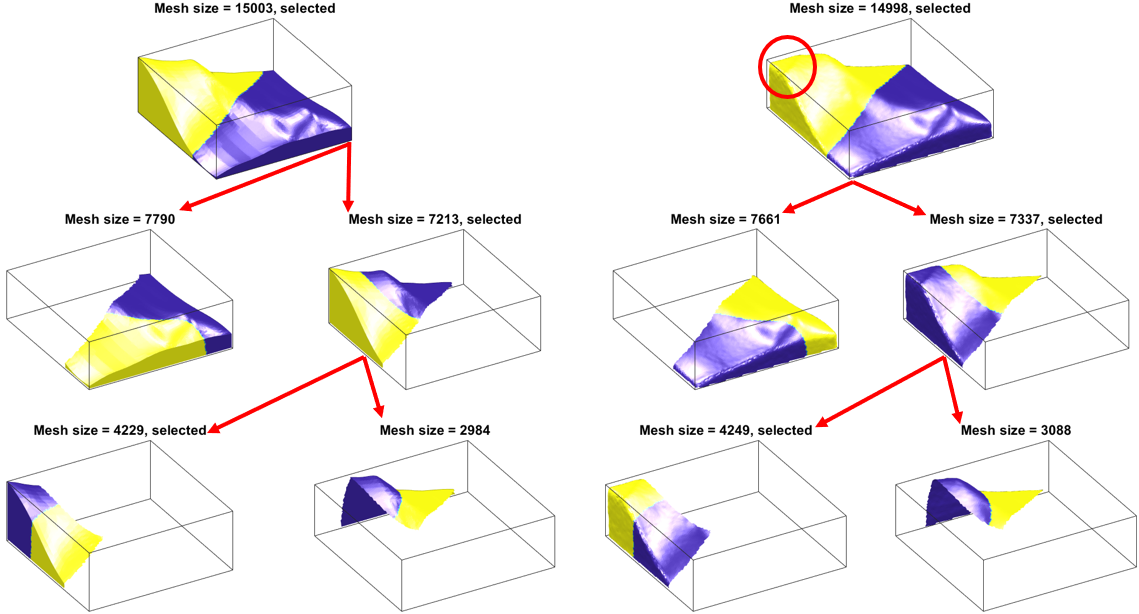}
    \caption{Recursive partitioning applied to a 3D-printed free-form piece to define a region of interest. Each row corresponds to an iteration, where starting from row 2, columns 1 to 4 represent sub-meshes $B+$, $B-$, $A+$, and $A-$, respectively. The red arrows indicate how the selected pair of components is further partitioned in the next iteration. Colors indicate the sign of the second LB eigenvector (yellow for positive and purple for negative). The algorithm correctly selects in the last iteration the submesh that contains the true local abnormality, circled in red on the top row, which is the ROI found.}
    \label{fig: ROI of Freeform}
\end{figure}

The proposed improved ROI algorithm is provided in Algorithm \ref{alg: ROI} shown in Appendix \ref{app 2}, which returns the ROI ${\mathcal{A}}_{ROI}$. Figures \ref{fig: ROI of Queen Chess Piece} to \ref{fig: ROI of Freeform} demonstrate the effectiveness of the algorithm in defining an ROI modified for the three parts used in prior sections.  After two iterations, the algorithm correctly identifies in each case the subregion where the abnormalities are located as the ROI. For readers wishing to see the details of the algorithm, a numerical step-by-step illustration as applied to the free-form surface is given in Figure \ref{fig:ROI_numeric} in Appendix \ref{app 2}.

\section{Advice to Practitioners}
\label{sec: summary}

In this paper, the practical implementation of the recently proposed spectral SPC methods for 3D geometrical data (\citeauthor{EDCZhao_QE} \citeyear{EDCZhao_QE}, \citeyear{zhaoEDC_Tech}, \citeyear{zhaoEDC_PE}, \citeyear{zhaoEDC_IJDS}) has been discussed, with Section \ref{sec: pre-processing} emphasizing the collection and preprocessing of meshes acquired from industrial-grade laser scanners and the issues that exist when working with real scanned meshes. We recommend practitioners follow the flowchart in Figure \ref{fig: PP_flowchart} with an emphasis on the necessity of the isotropic remeshing algorithm. Appendix \ref{app 1} showed the benefits of applying the isotropic remeshing algorithm to the meshes obtained by the scanner to improve numerical stability in the FEM computations of the Laplace-Beltrami operator, the main geometric feature used, and to promote near-orthogonality of the resulting eigenvectors. We recommend estimating the Laplace–Beltrami (LB) operator using the finite element method given. The previously open question of how to determine the best number of eigenvalues of the LB operator was answered in Section \ref{sec: No. Of Eig-vals} with a new approach based on the reconstruction of the CAD model of the part produced from its top eigenvectors. We recommend practitioners apply the proposed method to obtain the optimal number of eigenvalues to be used in the subsequent spectral SPC analysis. Section \ref{sec: SPC methods} demonstrated the reliability of both phase I and phase II methods, hence, we recommend the phase I control chart by \cite{capizzi2017phase} and the phase II control chart by \cite{chen2016distribution} with corrected moments to monitor process stability and detect process changes. When the ICP algorithm fails to register the CAD model with a part, we also recommend practitioners manually align the two meshes (e.g., using the $\tt{Align}$ function in Meshlab). In addition, a modification of the algorithm for defining a region of interest in \cite{zhaoEDC_IJDS} was presented in Section \ref{sec: ROI} with a new selection step based on all 4 pairwise distances between two submeshes from the nominal part and two submeshes from the part that triggered the online monitoring alarm. Contrary to the original algorithm, which has difficulties handling real scanned meshes, the new algorithm was demonstrated to converge to an ROI that includes the abnormality contained in all example parts we considered. Thus, we recommend performing the improved ROI algorithm to handle large scanned meshes efficiently.

While we concentrated on the implementation of the existing methods based on actual scanned meshes and proposed two methodological extensions, there are other areas where further research is necessary. The new method proposed for determining the best number of eigenvalues to monitor can indicate that, for a very complex part (e.g., with a lattice structure), many hundreds to thousands of eigenvalues are necessary to capture the geometrical shape of the object. A state-of-the-art multivariate nonparametric SPC method, such as \cite{capizzi2017phase}, which we used, has the constraint that the number of variables monitored must be less than the number of observations available during phase 1. This implies that the methods cannot be obtained for the phase 1 SPC monitoring of complex parts usually produced in small numbers. What is needed is a nonparametric, multivariate phase I SPC method that does not have this limitation. Furthermore, the LB spectrum is a ``profile" signal in nature, so developing new nonparametric phase I SPC methods that exploit the profile structure of the spectrum is of interest. Such a method should not require that the dimensionality of the profiles be smaller than the number of observations, given that large LB spectrum profiles could be necessary for parts with very complex geometry.

\subsection*{Data Availability Statement.} The supplementary materials contain computer code that implements all methods discussed and all the mesh datasets used in the examples, {\bf available in the JQT GitHub website:\\

\url{https://github.com/panr2000?tab=projects}}. \\

The \texttt{readme.docx} contains a step-by-step guide on how to use the main codes to aid practitioners in performing spectral SPC (phase I and phase II) with the tools reviewed in this paper. In addition, a description of all required packages, all data files, and all the included MATLAB code functions is given. 

\subsection*{Disclosure of interest.} The authors have no conflicts of interest to report.

\subsection*{Acknowledgments.} This work was funded by NSF grant CMMI 2121625. We thank Prof. Sanjay Joshi and Mr. Ashish Jacob (Penn State U.) for their assistance with the 3D printing of parts.

\appendix
\section{Isotropic Remeshing Algorithm (IRA)}
\label{app 1}
It is well known that in FEM computations, a large variability in the aspect ratio of the triangles in a mesh may result in numerical instabilities (see, e.g., \cite{FEMRef}). The aspect ratio of a triangle is simply the ratio of the lengths of the largest to the smallest edge. This implies that triangular meshes with a higher degree of {\em isotropy} should be preferred. The isotropy of a triangulation refers to the variance of the lengths of the edges of the triangles used. If all triangles tend to be of similar size and are equilateral, the triangulation has high isotropy (see Figure \ref{fig: Isotropy}). 

In all computations in this paper, the Isotropic Remeshing Algorithm (IRA) was applied to the scanned meshes as a preprocessing step \citep{alliez2003isotropic}. IRA aims to maximize the {\em isotropy} of a given mesh, a triangulation in our case, under certain specifications. The IRA transforms a mesh with low isotropy triangles into a mesh with high isotropy triangles, improving the numerical stability in the FEM computations \citep{botsch2010polygon}. 
\setcounter{figure}{0}
\begin{figure}[ht]
  \begin{subfigure}{0.45\textwidth}
  \centering
    \includegraphics[scale = 0.3]{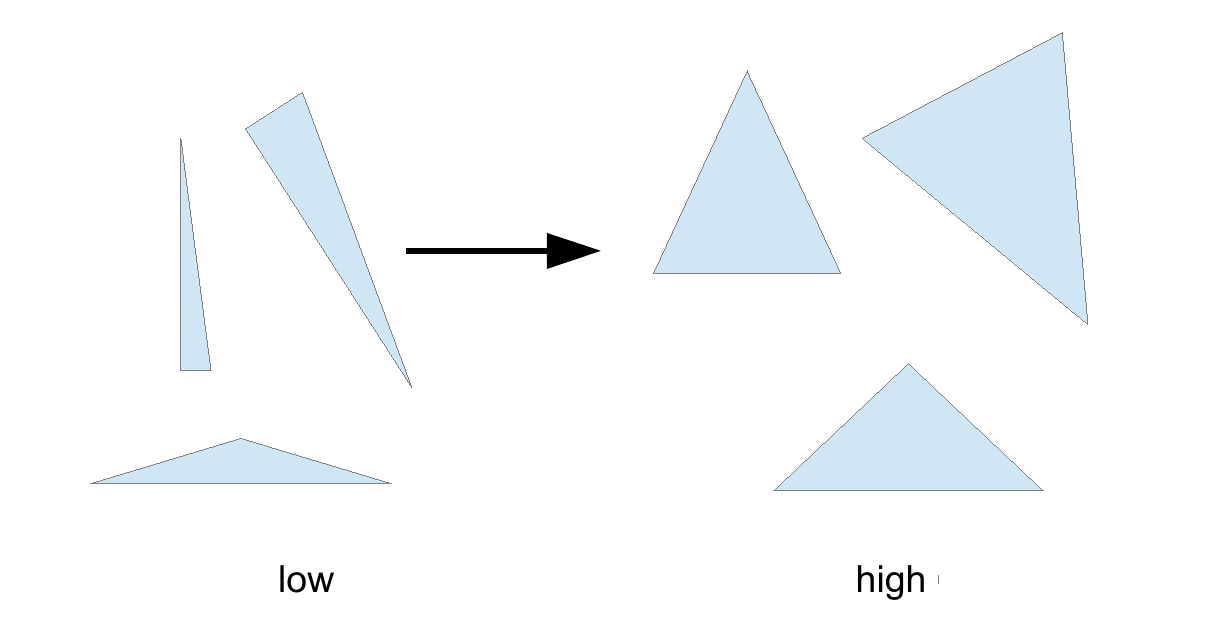}
  \end{subfigure}
  \begin{subfigure}{0.25\textwidth}
  \centering
    \includegraphics[scale = 0.175]{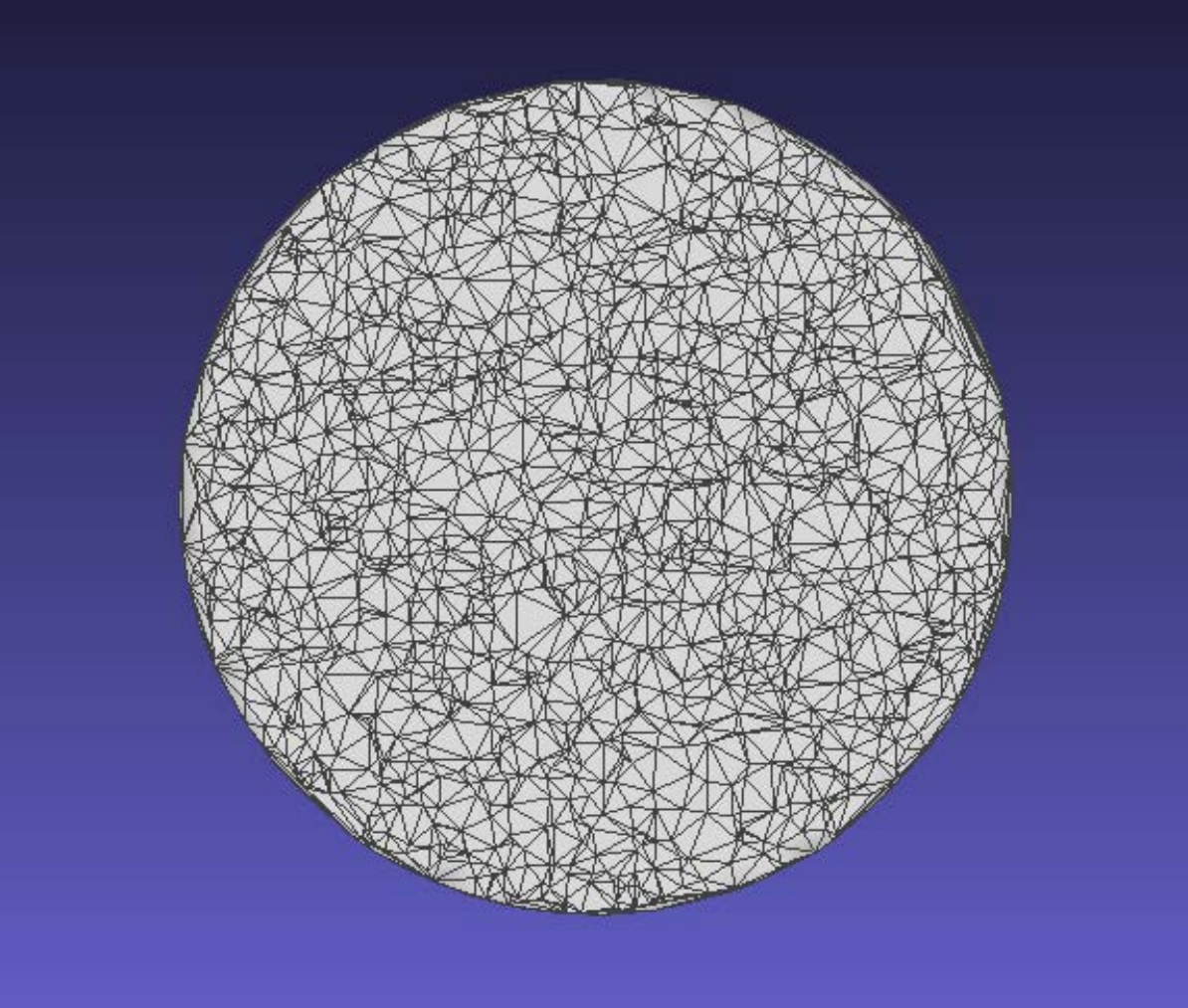}
  \end{subfigure}
  \begin{subfigure}{0.25\textwidth}
  \centering
    \includegraphics[scale = 0.175]{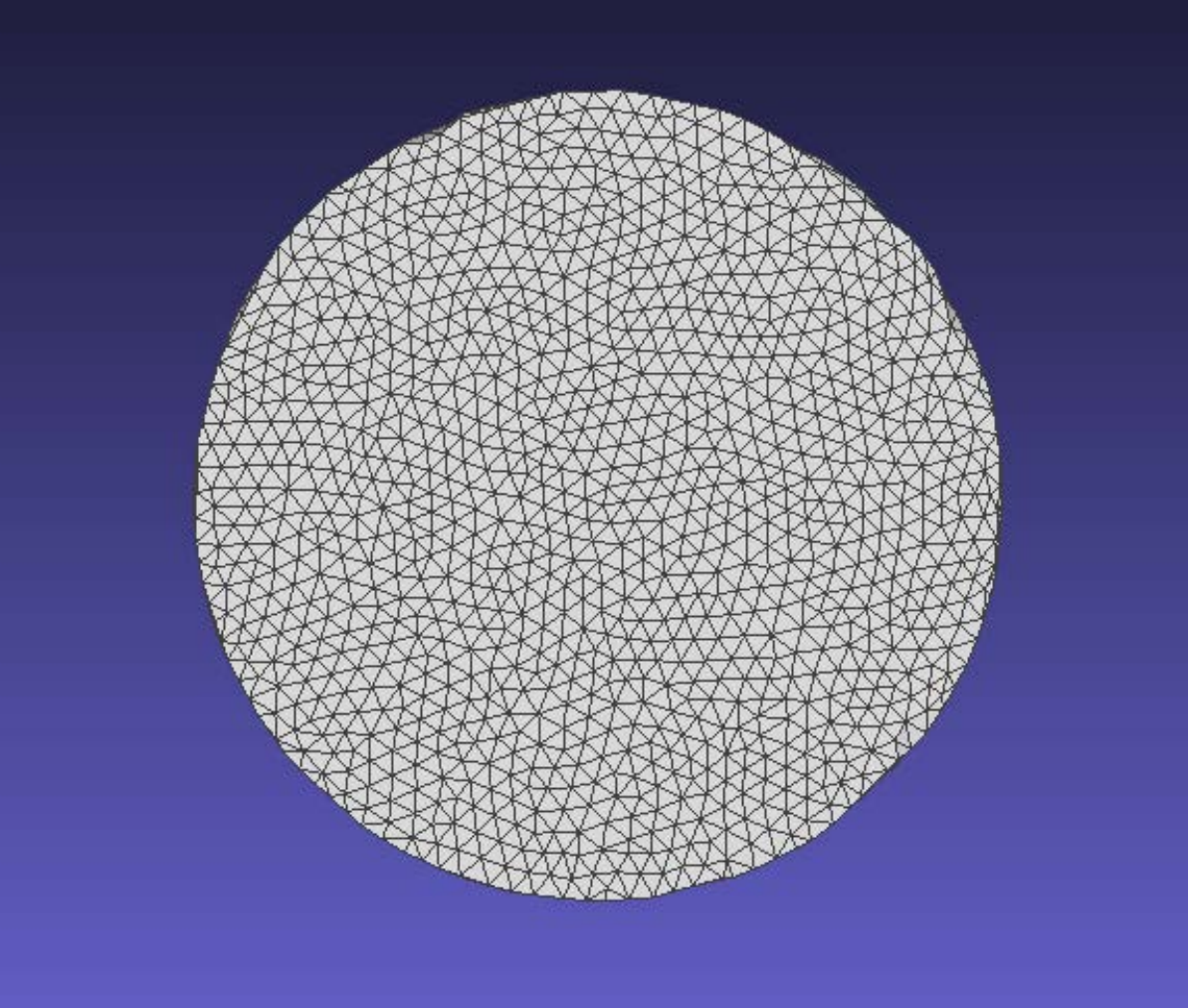}
  \end{subfigure}
  \caption{Left: low and higher isotropy triangles. Right: a mesh before and after applying the Isotropic Remeshing Algorithm (IRA)}
  \label{fig: Isotropy}
\end{figure}

A second justification for using IRA as preprocessing in the spectral SPC methods we use is numerical consistency. To illustrate, figure \ref{fig: Comparison} shows four different types of FEM estimators for the LB spectrum of a scanned mesh (obtained under different shape functions and different boundary conditions as discussed in \cite{zhaoEDC_PE})  of a commercial electrical box, computed with and without applying IRA remeshing as preprocessing. We can see that applying IRA to the scanned mesh improves the consistency of the estimated spectra regardless of the choice of the shape functions used by FEM or the boundary conditions utilized.
\setcounter{figure}{1}
\begin{figure}[ht]
\centering
  \begin{subfigure}{0.25\textwidth}
  \centering
    \includegraphics[scale = 0.4]{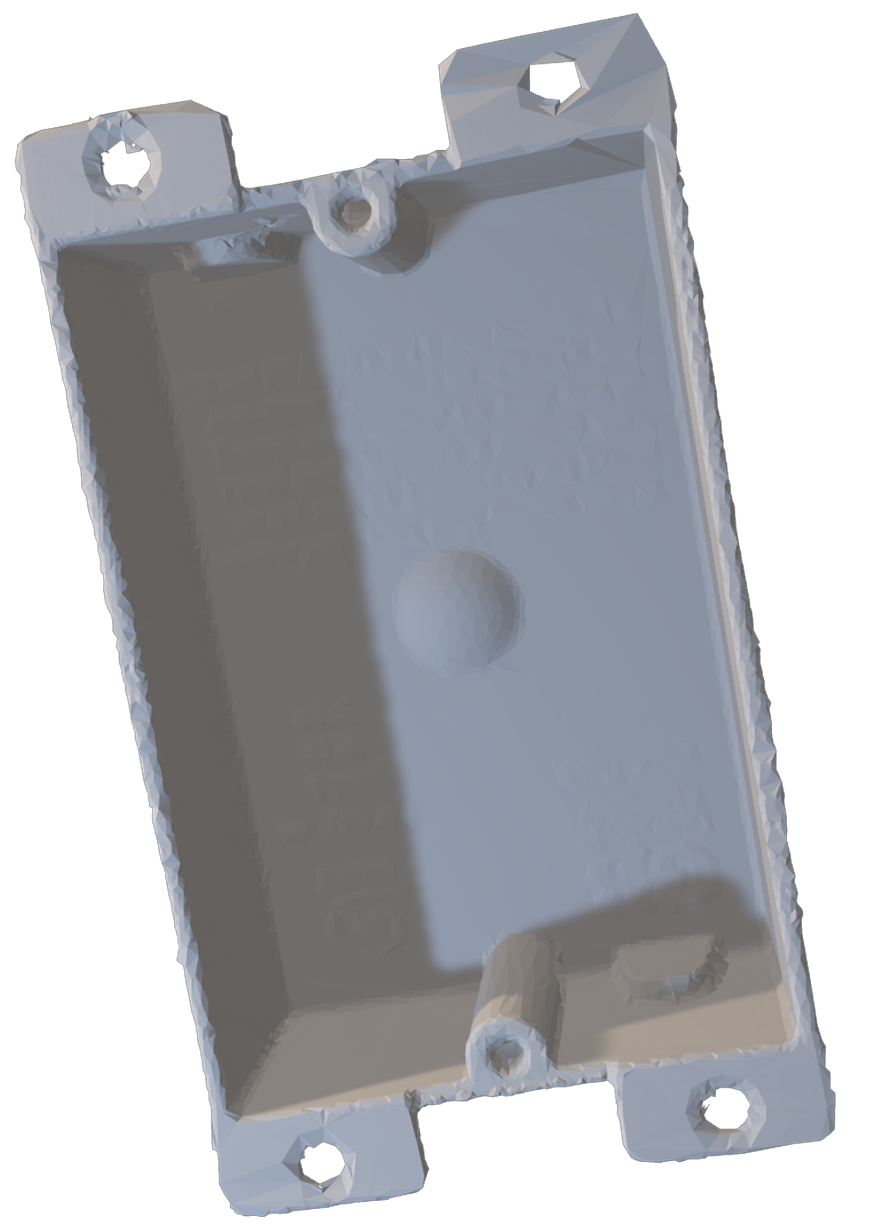}
  \end{subfigure}
  \begin{subfigure}{0.35\textwidth}
  \centering
    \includegraphics[scale = 0.3]{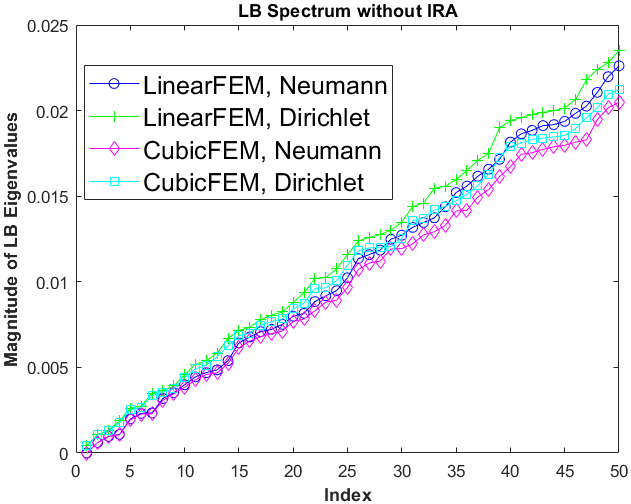}
    \end{subfigure}
  \begin{subfigure}{0.35\textwidth}
  \centering
    \includegraphics[scale = 0.3]{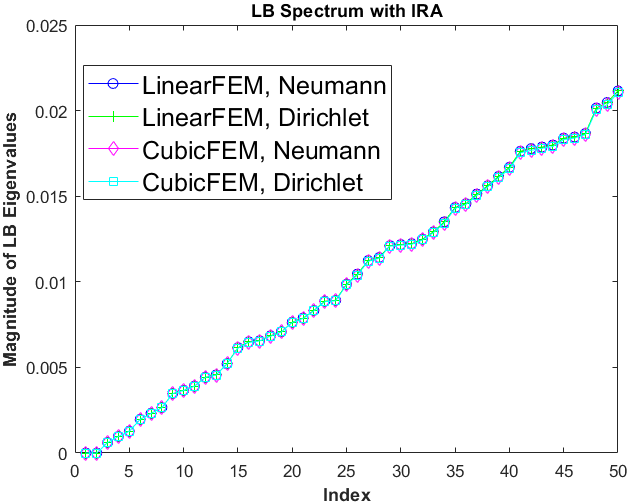}
  \end{subfigure}
  \caption{LB operator spectrum of the part on the left estimated without IRA (middle) and after IRA remeshing (right), using four variants of FEM estimators from \cite{zhaoEDC_PE}, obtained with different boundary conditions (Neumann or Dirichlet) and shape functions (linear or cubic).  With IRA remeshing, the different FEM variants get very close to each other, indicating the LB spectral computations have become more consistent compared to not applying any remeshing.}
  \label{fig: Comparison}
\end{figure}

Another matter of importance is that when applying a remeshing algorithm, the estimated LB spectra should be robust to noise. To investigate this, we compared the theoretical LB spectrum of a unit sphere with the FEM-approximated LB spectrum (with IRA) applied on noisy meshes simulated with different $\sigma-$noise levels. We use a unit sphere given that it is a continuous object for which the analytic or true LB operator spectrum is known \citep{reuter2006laplacebook}. Figure \ref{fig: comparison of 2-sphere}  shows a noise-free unit sphere to which different noise levels were added and  IRA was applied before the FEM computation. As can be seen from the figure, the FEM estimated LB spectrum under IRA is robust with respect to moderate levels of noise.
\setcounter{figure}{2}
\begin{figure}[ht]
    \begin{minipage}[t]{0.25\textwidth}
        \centering
        \begin{subfigure}[t]{\textwidth}
        \centering
            \includegraphics[width=0.6\linewidth]{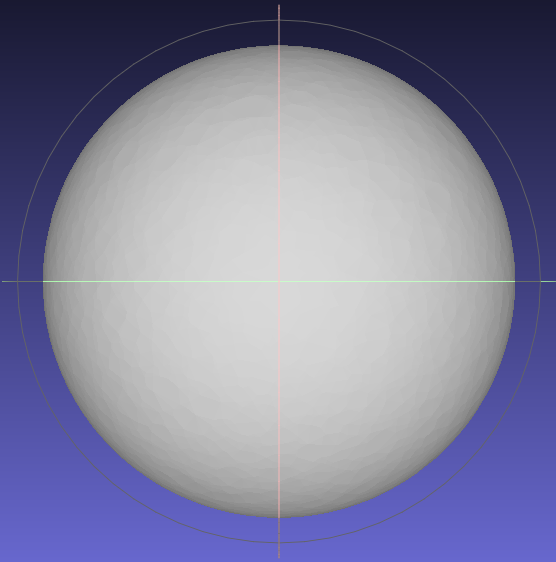}
            \caption{noise-free}
        \end{subfigure}
        \vfill
        \begin{subfigure}[t]{\textwidth}
            \centering
            \includegraphics[width=0.6\linewidth]{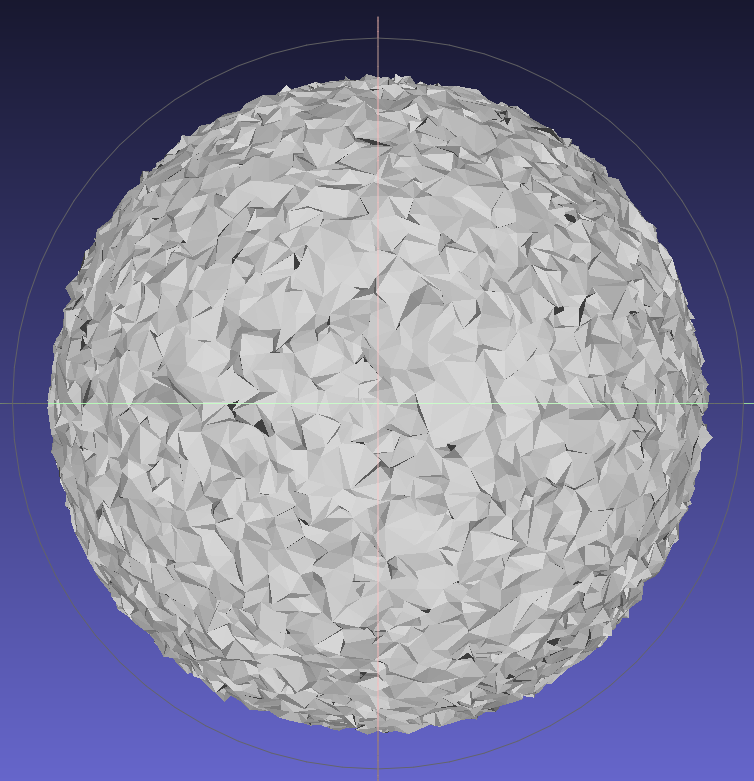}
            \caption{noise 0.01}
        \end{subfigure}
        \vfill
        \begin{subfigure}[t]{\textwidth}
        \centering
            \includegraphics[width=0.6\linewidth]{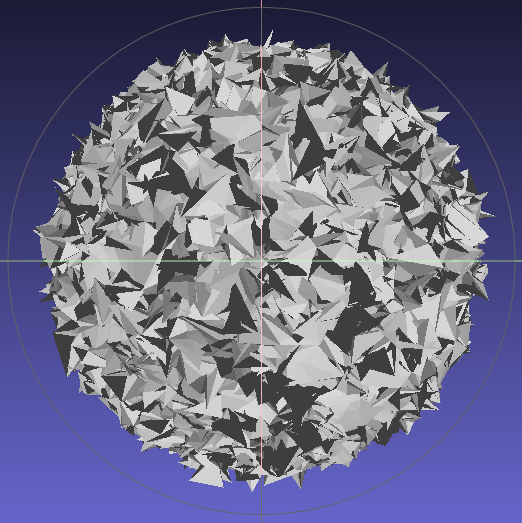}
            \caption{noise 0.05}
        \end{subfigure}
    \end{minipage}
    \begin{minipage}[t][\dimexpr\textheight-42\baselineskip][c]{0.65\textwidth}
        \centering
        \begin{subfigure}[c]{\textwidth}
        \centering
            \includegraphics[width=1.1\linewidth]{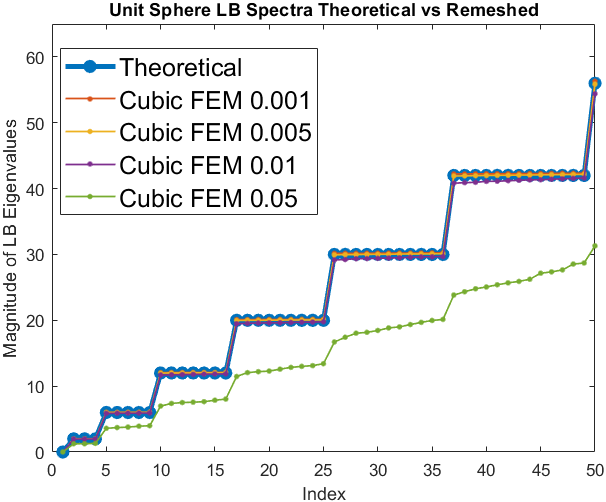}
        \end{subfigure}
    \end{minipage}
    \caption{Left panel:  Noise-free unit sphere,  and simulated spheres to which N(0,$\sigma^2)$ i.i.d. noise was added to each vertex. IRA remeshing was applied to all noisy meshes. Right panel: Known theoretical LB spectrum of noise-free unit sphere and FEM-estimated LB spectrum under different levels of noise $\sigma$. For moderate noise, the FEM-estimated LB spectrum deviates little from the theoretical spectrum, as expected. At $\sigma=0.05$ the surface geometry is far from a smooth sphere, and therefore the LB spectrum differs considerably more from the theoretical spectrum of a noise-free sphere.}
    \label{fig: comparison of 2-sphere}
\end{figure}

In addition to numerical stability, consistency, and robustness to noise of the estimated FEM LB spectra, we also require that the remeshing does not alter the geometrical shape of the given scanned mesh. To further study how IRA preserves the geometry of the scanned surface mesh data, we registered the noise-free CAD model, the scanned mesh, and the postprocessed scanned mesh with IRA for a commercial electrical box object, and computed their point-to-point deviations. Figure \ref{fig: p2p deviation} shows color-coded point-to-point deviations between the CAD model, the original scanned mesh, and the postprocessed scanned mesh using IRA. Darker color corresponds to smaller deviations, while lighter color corresponds to larger deviations. We can see from the figure how the deviations between the original scan and the IRA postprocessed mesh (second panel) are darker (lower) than the deviations from CAD to scanned mesh and from CAD to postprocessed mesh. Therefore, it can be concluded that IRA remeshing does not modify the surface geometry of the mesh data acquired by the scanner.
\setcounter{figure}{3}
\begin{figure}[ht]
\centering
  \begin{subfigure}{0.3\textwidth}
  \centering
    \includegraphics[scale = 0.4]{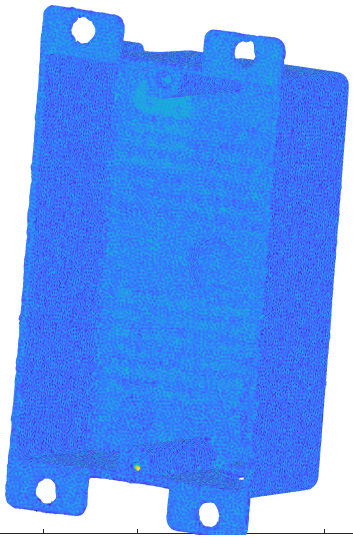}
    \caption{CAD vs Scan}
  \end{subfigure}
  \begin{subfigure}{0.3\textwidth}
  \centering
    \includegraphics[scale = 0.4]{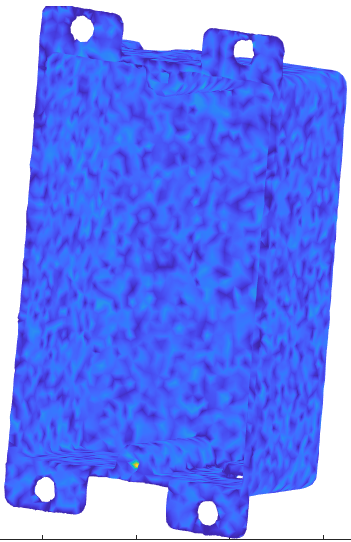}
    \caption{Scan vs PostProcessed}
  \end{subfigure}
  \begin{subfigure}{0.3\textwidth}
  \centering
    \includegraphics[scale = 0.4]{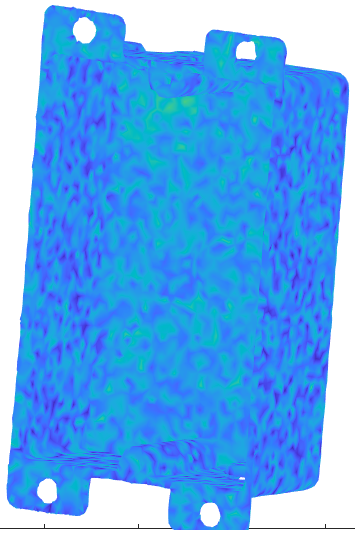}
    \caption{CAD vs PostProcessed}
  \end{subfigure}
  \caption{Point-to-point deviations between the mesh of the CAD model (633k vertices), scanned mesh (163k vertices), and IRA applied on the scanned mesh (14k vertices). The dark color corresponds to a small deviation, while the light color corresponds to a large deviation.}
  \label{fig: p2p deviation}
\end{figure}

A final, more subtle justification for using IRA remeshing comes from the desirability of obtaining orthogonal eigenvectors when solving the generalized eigenvalue problem. Since the Laplace-Beltrami operator in the continuous surface is a Hermitian, or self-adjoint operator, its eigenfunctions associated with different eigenvalues are orthogonal and form an orthogonal basis in $\mc{M}$, i.e., $\langle \phi_i,\phi_j\rangle_{\mathcal M}=0$ for $i \neq j$ \citep{belkin2008towards}. However, the eigenvectors of the discrete LB operator approximations do not form an orthogonal basis of functions if using the usual inner product in Euclidean space \citep{rustamov2007laplace}. That is, $\langle \phi_i, \phi_j\rangle = \phi_i^T \phi_j$ is {\em not} zero. Given the Fourier analysis interpretation of the eigendecomposition of the LB operator, it is desirable to retain this orthogonality in the eigenvectors once the operator is discretized. The orthogonality of the eigenvectors is used for reconstructing the object from its eigenvectors (see next section), to establish a functional map between two surfaces \citep{zhaoEDC_IJDS}, and to filter the noise of a random function defined on a surface \citep{Bronstein2017}.

\cite{ValletLevy2007} discuss how the orthogonality of the eigenfunctions of the continuous operator corresponds when using FEM, to the orthogonality of the eigenvectors under the $\mathbf B$ inner space, i..e, $\langle \phi_i, \phi_j\rangle_B = \phi_i^T \mathbf{B} \phi_j = 0$. Strict orthogonality is therefore not possible in general because $\mathbf{B} \neq \bf{I}$. Two additional results are useful here.
First, \cite{ValletLevy2007} show also how the FEM-estimated LB operator obtained by solving the generalized eigenvalue problem is equivalent to the so-called cotangent method for obtaining the LB estimator, an earlier approach in computing the LB operator in {\em closed} meshes, used in computer graphics \citep{Bronstein2017}. Then, \cite{rustamov2007laplace} shows that the eigenvectors from the cotangent method are orthogonal with respect to the $\mathbf{S}$ inner product, namely $\langle \phi_i,\phi_j \rangle _S = \phi_i^T \mathbf{S} \phi_j =0$, where
$\mathbf{S}$ is a diagonal matrix containing the areas of each of the triangles in the triangulation. It then follows that if all the triangles in a mesh had the same area $s$, it would imply $s\phi^T\phi_j=0$ or the orthogonality under the usual inner product in Euclidean space. Therefore, the conclusion is that a remeshing algorithm such as IRA, which reduces the variability of the triangle areas is desirable for the estimation of the LB operator spectrum via FEM since it will make its eigenvectors closer to orthogonal.

\section{Reconstruction of a mesh from top $k$ LB operator eigenvectors}
\label{app_Reconstruct}

To perform the CAD mesh reconstruction given its first $k$ eigenvectors, we need an orthogonal basis defined in $\mc{M}$ \citep{zhaoEDC_IJDS}. Given that the eigenvectors of the discrete LB operator do not form an orthogonal basis in $\mathcal{M}$ (even after IRA remeshing), we can apply the Gram-Schmidt orthogonalization process to the eigenvectors to obtain an orthonormal basis which we will denote $\{u_i\}_{i=1}^N$. For a given number of eigenvectors, $k$, with $1< k \leq N$, we denote by $U_k$  the matrix containing column vectors $\{u_i\}_{i=1}^k$, i.e.,
\[U_k =  \begin{bmatrix}
    \mid & \cdots & \mid \\
    u_1 & \cdots & u_k \\
    \mid & \cdots & \mid
\end{bmatrix}.\]
Define by $P_0$ the $N \times 3$ matrix of vertex coordinates of the CAD mesh model. Then the  matrix of coordinates of the reconstructed mesh, $P_k$, can simply be obtained by the projection of the CAD model matrix of coordinates onto the subspace spanned by the orthonormal columns of $U_k$, that is
\[P_{k} =Proj_{_{U_k}}(P_0) = U_k ( U_k^T U_k)^{-1} U_k^T P_0 = U_k U_k^T P_0,\]
since $( U_k^T U_k)^{-1} = I_k$. Note that $P_0 = P_N$ as the entire mesh is completely reconstructed when using all $N$ orthogonalized eigenvectors as they provide an orthonormal basis in $\mc{M}$.

\section{Improved Algorithm for Defining a Region of Interest}
\label{app 2}

The method for finding an ROI is based on Courant’s nodal domain theorem \citep{Buser}. Define the nodal lines of a surface (or manifold, in general) $\mc{M}$ as the set of points $x$ that satisfy $\phi_k(x) = 0$, i.e., these lines are formed by the roots of the eigenfunctions of its Laplace-Beltrami operator. Courant’s nodal domain theorem states that the nodal lines of the $k$-th eigenfunction of the LB operator divide a manifold $\mc{M}$ in no more than $k$ subregions, called nodal domains (see \cite{chavel1984eigenvalues}). Recall that from graph theory, the algebraic multiplicity of $\lambda_1=0$, the first graph Laplacian eigenvalue, equals the number of connected components in the graph \citep{Chung1997}. Thus, for instance, in a scanned mesh from a part, assuming they have a single connected component, the second eigenfunction, corresponding to the first non-zero eigenvalue, partitions the mesh into no more than 2 submeshes.

\begin{figure}[ht]
    \centering
    \includegraphics[scale = 0.4]{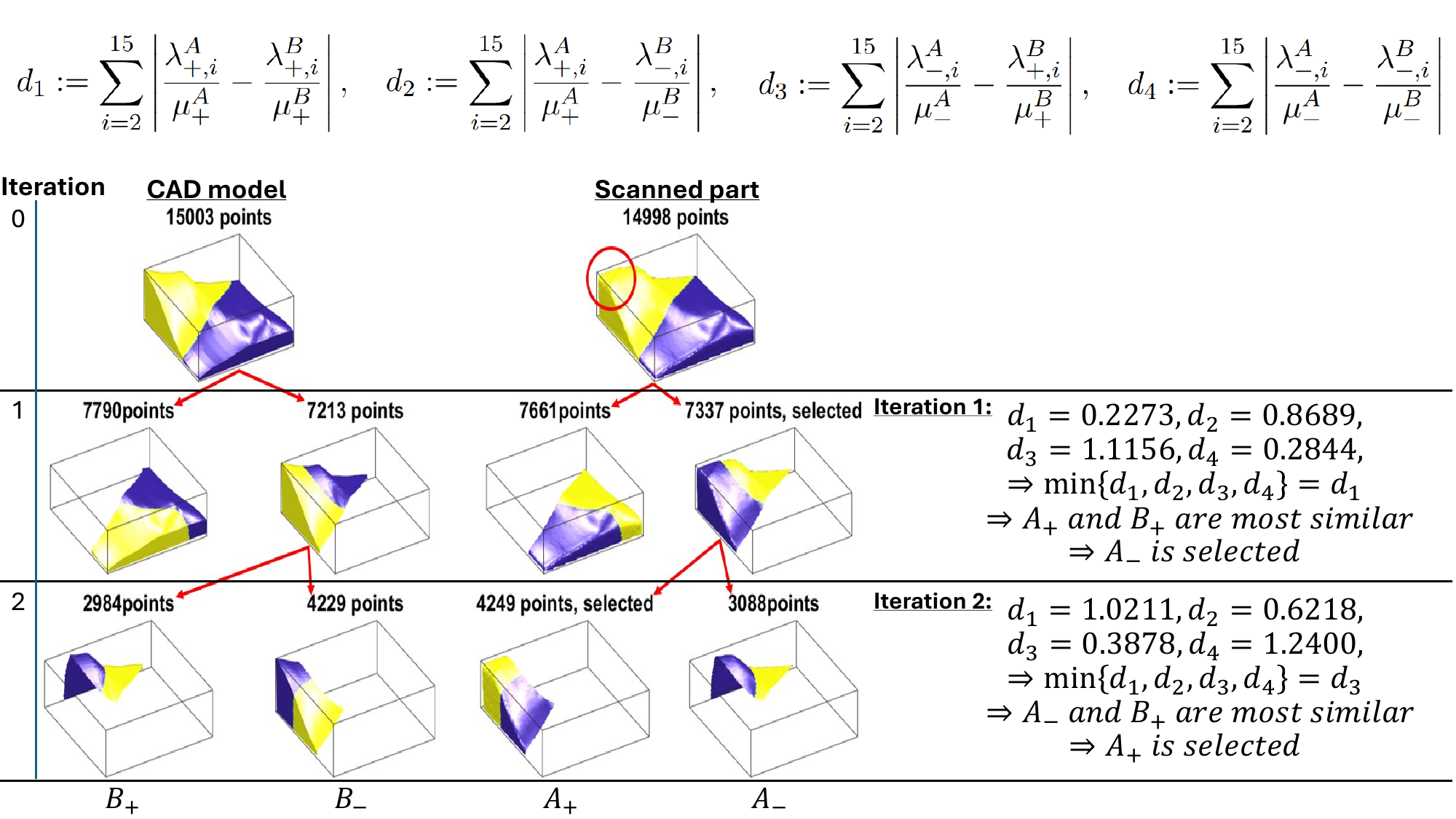}
    \caption{Numerical step-by-step illustration of the ROI-finding algorithm as applied to the free-from surface part. The scaled differences between the top $k$ (15 in this case) eigenvalues are listed in each of the two iterations, after which submesh $\mathcal{A}_{+}$ is considered small enough  (4249 vertices) and is selected as the ROI on the scanned part. Note this ROI contains the abnormality highlighted in Figure \ref{fig: Phase II SPC Freeform}. }
    \label{fig:ROI_numeric}
\end{figure}

Shown below is the algorithm, which takes the noise-free CAD model and the scanned part whose ROI one desires. Figure \ref{fig:ROI_numeric} provides a numeric illustration of the computational steps for one of the 3D free-form surfaces.

\begin{algorithm}[ht]
\caption{Improved Recursive Method to Define a Region of Interest. Modified from \cite{zhaoEDC_IJDS}.}\label{alg: ROI}
\textbf{Input:} Abnormal part mesh $\mathcal{A}$, CAD mesh $\mathcal{B}$, number of eigenvalues $k$. \\
\textbf{Output:} A submesh of on $\mathcal{A}$ representing a region of interest $\mathcal{A}_{ROI}$  
\begin{algorithmic}
\For{i=1:iter}
    \State  Partition $\mathcal{A}$ into $\mathcal{A}_{+}$  and  $\mathcal{A}_{-}$ based on the sign of the second LB 
    \State eigenvector, estimated using linear FEM 
    \State  Partition $\mathcal{B}$ into $\mathcal{B}_{+}$  and  $\mathcal{B}_{-}$ based on the sign of the second LB 
    \State eigenvector, estimated using linear FEM
\State  Calculate the first $k$ eigenvalues for $\mathcal{A}_{+}, \mathcal{A}_{-}, \mathcal{B}_{+}, \mathcal{B}_{-}$, respectively
\State  Calculate $d_1,d_2,d_3,d_4$
\If{$d_1 = min\{d_1,d_2,d_3,d_4\}$}
    \State Set $\mathcal{A} = \mathcal{A}_-$, and $\mathcal{B} = \mathcal{B}_-$
  \ElsIf{$d_2 = min\{d_1,d_2,d_3,d_4\}$}
    \State  Set $\mathcal{A} = \mathcal{A}_-$, and $\mathcal{B} = \mathcal{B}_+$
  \ElsIf{$d_3 = min\{d_1,d_2,d_3,d_4\}$}
    \State  Set $\mathcal{A} = \mathcal{A}_+$, and $\mathcal{B} = \mathcal{B}_-$
  \Else
    \State  Set $\mathcal{A} = \mathcal{A}_+$, and $\mathcal{B} = \mathcal{B}_+$
  \EndIf{}
  \EndFor
  \State Return $\mathcal{A}_{ROI} = \mathcal{A}$
\end{algorithmic}
\end{algorithm}

\end{document}